\documentclass[prd,aps,a4paper,twocolumn,eqsecnum,nofootinbib,floatfix]{revtex4}  

\newif\ifusesec
\usesectrue  
 
\usepackage{longtable}
\usepackage{graphicx} 
\usepackage{amsmath,amsfonts,amssymb}

\newcommand{\beq}{\begin{equation}}
\newcommand{\eeq}{\end{equation}}
\newcommand{\bea}{\begin{eqnarray}}
\newcommand{\eea}{\end{eqnarray}}
\newcommand{\ch}{C}
\newcommand{\sh}{S}

\begin{document}

\title{Frequency domain analysis of the gravitational wave energy loss in hyperbolic encounters}

\author{Donato Bini$^{1,2}$, Andrea Geralico$^1$}
  \affiliation{
$^1$Istituto per le Applicazioni del Calcolo ``M. Picone,'' CNR, I-00185 Rome, Italy\\
$^2$INFN, Sezione di Roma Tre, I-00146 Rome, Italy\\
}

\date{\today}

\begin{abstract}
The   energy radiated (without the 1.5PN tail contribution which requires a different treatment) by a binary system of compact objects moving in a hyperboliclike orbit is computed in the frequency domain through the second post-Newtonian level as an expansion in the large-eccentricity parameter    
up to next-to-next-to-leading order, completing the time domain corresponding information (fully known in closed form at the second post-Newtonian of accuracy). 
The spectrum contains quadratic products of the modified Bessel functions of the first kind (Bessel K functions) with frequency-dependent order (and argument) already at Newtonian level, so preventing the direct evaluation of Fourier integrals.
However, as the order of the Bessel functions tends to zero for large eccentricities, a large-eccentricity expansion of the spectrum allows for analytical computation beyond the lowest order.
\end{abstract}

\maketitle

\section{Introduction}

The orbital-averaged gravitational-wave energy and angular momentum fluxes as well as the secular evolution of orbital elements under gravitational radiation reaction are the basic input for constructing templates for the binary dynamics.
Most of the attention has been devoted to ellipticlike orbits, starting from the pioneering works of Peters and Mathews \cite{Peters:1963ux,Peters:1964zz}, who computed both the average energy and angular momentum emission rates at Newtonian order.
Their results have been then improved over the years up to the third Post-Newtonian (PN) level \cite{Wagoner:1976am,Damour:1988mr,Blanchet:1989cu,Junker:1992kle,Schaefer:1993,Blanchet:1995fg,Blanchet:1995ez, Gopakumar:1997bs,Blanchet:2004ek, Arun:2007sg}, also including nonlinear effects of tails starting at 1.5PN \cite{Blanchet:1993ec}. 
The averaging procedure is usually done in the time domain by using a quasi-Keplerian representation of the motion. However, the computation in the frequency domain proves useful too, especially when dealing with tail contributions.
In fact, in this case the multipole moments from which the fluxes are constructed have a more complicated dependence on time, preventing the integrals to be analytically evaluated in simple closed forms \cite{Arun:2007rg}.
The dependence on the eccentricity is generally encoded in a number of so-called \lq\lq enhancement factors," which enhance the effect with respect to the quasi-circular case.
Most of them do not admit  closed-form expressions, and are determined by numerical fits. 
This is the case of tail integrals, which have been evaluated in Refs. \cite{Blanchet:1993ec,Arun:2007rg} in the form of infinite series involving quadratic products of Bessel functions of the first kind (Bessel $J$ functions).
However, it is always possible to obtain analytical expressions as power series in the small eccentricity.

Peters also considered the hyperbolic scattering of a small mass by a large mass \cite{Peters:1970mx}, leading to the emission of the so-called gravitational bremsstrahlung radiation. In fact, in close analogy with the electromagnetic case of an electron scattered by a nucleus, the small mass is expected to emit a burst 
of multipole radiation as it experience a sudden transverse acceleration due to the large mass as it flies by (see also Refs. \cite{rr1,rr2}). Peters then studied the spectrum and angular distribution of the radiation in the limiting situations of both slow and high velocities, for large values of the impact parameter.
This problem has been further investigated by many authors by using different methods and approximations (see, e.g., Refs. \cite{Kovacs:1977uw,Kovacs:1978eu} and references therein), leading to different results in the ultrarelativistic regime (see also Refs. \cite{Bern:2021dqo,Herrmann:2021lqe} for recent developments).

The aim of the present work is to compute the total instantaneous energy radiated during a hyperbolic encounter in the frequency domain through the 2PN level, without including hereditary (tail, tail-of-tail and tail-squared) terms. The tail-of-tail and tail-squared contributions will be deeply discussed in a forthcoming paper \cite{bin_ger_to_appear}, whereas the 1.5PN (linear) tail contributions have been recently studied in Ref. \cite{Bini:2021gat}.

This is the current knowledge of the averaged energy flux recently obtained in Ref. \cite{Bini:2020hmy} working but in the time domain, generalizing previous results \cite{Turner:1977,Blanchet:1989cu}. In the literature, the Fourier analysis has been adopted at the Newtonian level only following early studies \cite{rr1,rr2}, and at the leading order in a large impact parameter expansion (splash radiation). In this limit the power spectrum is given in terms of quadratic products of modified Bessel functions of the first kind (Bessel $K$) with integer order, which allow for analytical integration. Releasing the assumption of large impact parameter (equivalent to large eccentricity if a quasi-Keplerian parametrization of the orbit is used) implies that the order of the Bessel functions be frequency-dependent, so preventing the Fourier integrals to be computed in closed form by using standard techniques \cite{Zeldovich:1974}. 

A similar situation occurs in the case of electromagnetic dipole radiation emitted by a system of two attracting charges in relative hyperbolic motion \cite{Landau:1982dva}.
However, the order of the Bessel functions goes to zero for large eccentricities, so that one can take a large-eccentricity expansion of the power spectrum allowing for analytical computation beyond the leading order \cite{Bini:2017wfr}.
We will show here how is possible to go beyond this level, by computing the averaged energy flux up to the next-to-next-to-leading order (NNLO). 
We will provide all technical details to go even further in this expansion by a systematic use of the Mellin transform (following a procedure already sketched in Ref. \cite{Bini:2020rzn}), so obtaining the hyperbolic counterparts of the enhancement functions for the elliptic case as series in inverse powers of the large eccentricity.

Following standard notations, we define the symmetric mass ratio $\nu\equiv \frac{\mu}{M}$ as the ratio of the reduced mass $\mu\equiv m_1 m_2/(m_1+m_2)$ to the total mass $M=m_1+m_2$. 
Other standard notations will be used, like $X_1=\frac{m_1}{M}$ and $X_2=\frac{m_2}{M}$, with $X_1+X_2=1$.
We shall work with dimensionless variables $r=c^2 r^{\rm phys}/(GM)$, $t= c^3 t^{\rm phys}/(GM)$ as well as dimensionless rescaled orbital parameters, e.g., $a_r \equiv c^2 a^{\rm phys}/( GM)$ for the semi-major axis.
We generally use units where $c$ and $G$ (and sometimes also $GM$) are set to unity. 
However, we will keep track of the fractional PN order of the various contributions to PN-expanded quantities by using the placeholder $\eta \equiv \frac1c$, so that 
$\eta^0$ stands for the Newtonian level of accuracy,  $\eta^2$ for the 1PN, etc. 

\section{The harmonic-coordinate quasi-Keplerian parametrization of the hyperbolic  motion}
 
We will adopt the 2PN-accurate quasi-Keplerian description of the binary dynamics in harmonic coordinates \cite{DD1981a,D1982,dd,Damour:1988mr,Damour:1990jh,Schaefer:1993,Cho:2018upo}
\begin{eqnarray} \label{hypQK2PN}
r&=& \bar a_r (e_r \cosh v-1)\,,\nonumber\\
\bar n t&=&e_t \sinh v-v + f_t V+g_t \sin V\,,\nonumber\\
\phi &=&K[V+f_\phi \sin 2V+g_\phi \sin 3V]\,,
\end{eqnarray}
with
\beq
\label{Vdef}
V=2\, {\rm arctan}\left[\sqrt{\frac{e_\phi+1}{e_\phi-1}}\tanh \frac{v}{2}  \right]\,,
\eeq
where we have used dimensionless variables and $c=1$.
The expressions of the orbital parameters $\bar n$, $\bar a_r$, $K$, $e_t,e_r,e_\phi$, $f_t,g_t,f_\phi, g_\phi$ are given, e.g., in Table VIII of Ref. \cite{Bini:2020hmy} as functions of the dimensionless specific binding energy $\bar E \equiv (E_{\rm tot}-Mc^2)/(\mu c^2)$ and the dimensionless angular momentum $j=c J/(GM\mu)$ of the system.
Note that the quasi-Keplerian parametrization of the motion in the hyperboliclike case can be seen as an analytic continuation of the ellipticlike one only up to 1PN \cite{dd,Cho:2018upo}.

We will take below $e_r$ and $\bar a_r$ as fundamental variables (a choice of any other pair of orbital parameters being equivalent, like e.g. energy and angular momentum), in terms of which the remaining orbital parameters write as
\bea
\bar n&=&(\bar a_r)^{-3/2}\left[
1+\frac{\eta^2}{2\bar a_r}(9-\nu)\right.\nonumber\\
&&\left.
+\frac{\eta^4}{\bar a_r^2}\left(\frac{147}{8}-\frac{25}{8}\nu+\frac{3}{8}\nu^2-\frac{3}{2}\frac{7\nu-4}{e_r^2-1}\right)
\right]
\,,\nonumber\\
K&=&1+\frac{3\eta^2}{\bar a_r(e_r^2-1)}\nonumber\\
&&
+\frac{\eta^4}{\bar a_r^2(e_r^2-1)^2}\left(\frac34(2\nu-3)e_r^2-6\nu+\frac{21}{2}\right)
\,,\nonumber\\
e_t&=&e_r\left[
1+\frac{\eta^2}{\bar a_r}\left(-\frac32\nu+4\right)\right.\nonumber\\
&&\left.
+\frac{\eta^4}{\bar a_r^2}\left(16-\frac{67}{8}\nu+\frac{15}{8}\nu^2-\frac{7\nu-4}{e_r^2-1}\right)
\right]
\,,\nonumber\\
e_\phi&=&e_r\left[
1-\frac{\nu\eta^2}{2\bar a_r}\right.\nonumber\\
&&\left.
+\frac{\eta^4}{32\bar a_r^2}\left(\nu(15\nu-29)-\frac{160+357\nu-15\nu^2}{e_r^2-1}\right)
\right]
\,,\nonumber\\
f_t&=&\frac{3\eta^4}{2\bar a_r^2}\frac{5-2\nu}{\sqrt{e_r^2-1}}
\,,\nonumber\\
f_\phi&=&\frac{\eta^4}{8\bar a_r^2}\frac{e_r^2(1+19\nu-3\nu^2)}{(e_r^2-1)^2}
\,,\nonumber\\
g_t&=&\frac{\eta^4}{8\bar a_r^2}\frac{e_r\nu(15-\nu)}{\sqrt{e_r^2-1}}
\,,\nonumber\\
g_\phi&=&\frac{\eta^4}{32\bar a_r^2}\frac{e_r^3\nu(1-3\nu)}{(e_r^2-1)^2}
\,.
\eea
$\bar a_r$ and $e_r$ can be then reexpressed in terms of $\bar E$ and $j$ by using the relations
\bea \label{arer}
\bar a_r &=& \frac{1}{2\bar E}\left[1-\frac{1}{2}\bar E \eta^2 (-7+\nu)\right. \nonumber\\
&&\left.+\frac{1}{4}\bar E^2\eta^4 \left(1+\nu^2-8 \frac{(-4+7\nu)}{\bar E j^2}\right)\right] 
\,,\nonumber\\
e_r^2 &=& 1+2\bar E j^2+\bar E [5\bar Ej^2(\nu-3)+2\nu-12]\eta^2\nonumber\\
&&
+\frac{\bar E}{j^2}[(4\nu^2+80-45\nu)\bar E^2j^4\nonumber\\
&&+(\nu^2+74\nu+30)\bar Ej^2+56\nu-32]\eta^4
\,.
\eea

\section{Computation of the gravitational wave energy loss in the frequency domain}

The total energy radiated during a hyperbolic encounter is given by integrating the instantaneous flux ${\mathcal F}^{\rm GW}(t)$ of gravitational wave energy during the whole process, i.e.,
\beq
\Delta E^{\rm GW}=\int dt {\mathcal F}^{\rm GW}(t)\,.
\eeq
The fractionally 2PN-accurate expression for ${\mathcal F}^{\rm GW}(t)$ in terms of multipole moments~\cite{Blanchet:1985sp,Blanchet:1987wq,Blanchet:1989ki,Damour:1990ji,Blanchet:1998in,Poujade:2001ie} reads
\beq
{\mathcal F}^{\rm GW}(t)=\frac{G}{c^5} \left[{\mathcal F}^{\rm GW}_{{\rm I}_2}(t)+\eta^2{\mathcal F}^{\rm GW}_{{\rm I}_3, {\rm J}_2}(t)+\eta^4{\mathcal F}^{\rm GW}_{{\rm I}_4,{\rm J}_3}(t)\right]\,,
\eeq
where
\begin{eqnarray}
\label{flux2PNdef}
{\mathcal F}^{\rm GW}_{{\rm I}_2}(t)&=& \frac15 I_{ab}^{\rm (3)}(t) I_{ab}^{\rm (3)}(t) 
\,,\nonumber\\
{\mathcal F}^{\rm GW}_{{\rm I}_3, {\rm J}_2}(t)&=& \frac1{189 } I_{abc}^{\rm (4)}(t) I_{abc}^{\rm (4)}(t) +\frac{16}{45 } J_{ab}^{\rm (3)}(t) J_{ab}^{\rm (3)}(t)
\,,\nonumber\\
{\mathcal F}^{\rm GW}_{{\rm I}_4,{\rm J}_3}(t)&=& \frac{1}{9072}I_{abcd}^{\rm (5)}(t) I_{abcd}^{\rm (5)}(t) +\frac{1}{84}J_{abc}^{\rm (4)}(t) J_{abc}^{\rm (4)}(t)
\,,\nonumber\\
\end{eqnarray}
the superscript in parenthesis  denoting  repeated time derivatives.  
The 2PN harmonic-coordinate expressions for the multipole moments can be found in Ref. \cite{Arun:2007rg}.

The first step is to Fourier transform the multipole moments. For example,
\beq
\label{I_ab_fourier}
I_{ab}(t)=\int \frac{d\omega}{2\pi}e^{-i \omega t}\hat I_{ab}(\omega)\,,
\eeq
where
\beq
\label{I_ab_omega}
\hat I_{ab}(\omega)=\int_{-\infty}^{+\infty} dt e^{i\omega t}I_{ab}(t)\,,
\eeq
with the associated PN expansion 
\bea
\label{PN_Iab}
\hat I_{ab}(\omega)&=&\hat I_{ab}^{\rm N}(\omega)+\eta^2\hat I_{ab}^{\rm 1PN}(\omega)\nonumber\\
&+&\eta^4\hat I_{ab}^{\rm 2PN}(\omega)
+O(\eta^6)\,.
\eea
For instance, at the Newtonian level we get
\bea
\Delta E_{\rm GW}^{\rm N}&=&\frac{G}{5 c^5}\int\, dt I_{ab}^{\rm N(3)}(t) I_{ab}^{\rm N(3)}(t) \nonumber\\
&=&\frac{G}{5 c^5} \int \frac{d\omega}{2\pi} \frac{d\omega'}{2\pi}(-i\omega)^3(-i\omega')^3\hat I^{\rm N}_{ab}(\omega)\hat I^{\rm N}_{ab}(\omega') \nonumber\\
&&
\times\,\int dt\,  e^{-i (\omega+\omega') t}
\nonumber\\
&=&\frac{G}{5 c^5} \int_{-\infty}^{+\infty} \frac{d\omega}{2\pi}\,\omega^6|\hat I^{\rm N}_{ab}(\omega)|^2
\,,
\eea
having used $\int_{-\infty}^{+\infty} dt\,  e^{-i (\omega+\omega') t}=2\pi\delta(\omega+\omega')$ and $\hat I^{\rm N}_{ab}(-\omega)=\hat I^{\rm N}_{ab}{}^*(\omega)$, the quadrupole moment being computed at the lowest order.
At 2PN we have
\beq 
\label{EGWomega}
\Delta E_{\rm GW}=\frac{G}{\pi c^5}\int_{0}^\infty  d\omega  {\mathcal K}(\omega)\,,
\eeq
where
\begin{eqnarray}
{\mathcal K}(\omega)&=&\frac15 \omega^6 |\hat I_{ab}(\omega)|^2\nonumber\\
&+&\eta^2 \left[\frac{\omega^8}{189}|\hat I_{abc}(\omega)|^2+\frac{16}{45}\omega^6 |\hat J_{ab}(\omega)|^2  \right]\nonumber\\
&+&\eta^4 \left[\frac{\omega^{10}}{9072}|\hat I_{abcd}(\omega)|^2+\frac{\omega^8}{84}|\hat J_{abc}(\omega)|^2  \right]\,,\nonumber\\
\end{eqnarray}
and each multipole moment is computed at the needed PN order necessary to reach the desired accuracy.

It is convenient to replace the integration over the frequency $\omega$ by an integration over the rescaled frequency variable $u$, using
\beq
\omega= \frac{u}{e_r \bar a_r^{3/2}} \,,
\eeq
so that Eq. \eqref{EGWomega} becomes
\beq \label{EGWu}
\Delta E_{\rm GW}=\frac{G}{\pi c^5}\frac{1}{e_r \bar a_r^{3/2}}  \int_{0}^\infty  du  {\mathcal K}(u)\,,
\eeq
with
\beq
{\mathcal K}(u)={\mathcal K}(\omega)\big|_{\omega= u/(e_r \bar a_r^{3/2})}\,.
\eeq
which can then be decomposed as 
\beq
\label{calKPNexp}
\mathcal K(u)=\mathcal K_{\rm N}(u)+\frac{\eta^2}{\bar a_r}\mathcal K_{\rm 1PN}(u)+\frac{\eta^4}{\bar a_r^2}\mathcal K_{\rm 2PN}(u)\,,
\eeq
where we used a combined  PM-PN expansion. The coefficients $\mathcal K_{\rm nPN}(u)$ will be given below in the large-eccentricity limit allowing for the explicit computation of the related energy integrals. 
The large-eccentricity expansion can be avoided when working in the time domain  where one can obtain easily 
a closed-form expression for the corresponding radiated energy.

\subsection{Fourier transform of the multipole moments}

We have to compute first
\beq
\label{I_ab_omega2}
\hat I_{ab}(\omega)=\int \frac{dt}{dv}e^{i\omega t(v)}I_{ab}(t)|_{t=t(v)} \, dv \,,
\eeq
and similarly for the other moments.

The PN expansion of the exponential term $e^{i\omega t(v)}$ gives
\beq
e^{i\omega t(v)}=e^{q\sinh v -p v}\left(1+\tilde b_2\frac{\eta^2}{\bar a_r}+\tilde b_4 \frac{\eta^4}{\bar a_r^2}\right)\,,
\eeq
where
\beq \label{defupq}
 u\equiv \omega e_r \bar a_r^{3/2}\,,\qquad q \equiv i \, u\,,\qquad  p \equiv \frac{q}{e_r }\,, 
\eeq 
so that $e_r=q/p$, and $\tilde b_2$ and $\tilde b_4$ do not depend on $\bar a_r$ but only on $v$, $p$ and $q$ and are given by
\bea
\tilde b_2&=&-\frac{1 }{2 } \left[(2\nu+1) q\sinh v +(\nu  -9)p v 
\right] \,,\nonumber\\
\tilde b_4-\frac12 \tilde b_2^2&=&\frac{1}{8} \left[\tilde b_{40}+\tilde b_{41} {\rm arctan}\left(\sqrt{\frac{e_r+1}{e_r-1}}\tanh \frac{v}{2}  \right) \right]
\,,\nonumber\\
\eea
with
\bea
\tilde b_{40}&=& q(8\nu^2-8\nu-1)\sinh v +pv (\nu^2-63+95\nu) \nonumber\\
&&+2 q (7\nu-4)\left[\frac{\sinh v -3 v}{e_r-1} -\frac{\sinh v+3 v}{e_r+1} \right]\nonumber\\
&&-q \frac{\nu( \nu-15)\sinh v}{e_r\cosh v-1} 
\,,\nonumber\\
\tilde b_{41}&=& -\frac{48}{ e_r \sqrt{e_r^2-1}}\left(\nu-\frac{5}{2}\right) \,.
\eea
Moreover,
\beq
\frac{dt}{dv}=\bar a_r^{3/2} (e_r\cosh v-1)\left(1 +\tilde c_2\frac{\eta^2}{\bar a_r} +\tilde c_4\frac{\eta^4}{\bar a_r^2}\right)\,,
\eeq
with 
\bea
\tilde c_2&=&-\frac1{2(e_r\cosh v-1)}   [(2\nu+1)e_r\cosh v +\nu-9] 
\,,\nonumber\\
\tilde c_4&=&\left(\nu^2-\nu-\frac18 \right) -\frac{\nu (-15+\nu)  (e_r^2-1)}{8(e_r\cosh v-1)^3} \nonumber\\
&&+\frac{(60-39\nu+\nu^2)}{8(e_r\cosh v-1)^2} +\frac{(9\nu^2+3\nu-16)}{8(e_r\cosh v-1)} \nonumber\\
&&  +\frac{(7\nu-4)}{(e_r^2-1)}\left(\frac12 -\frac{1}{(e_r\cosh v-1)}\right)
\,.
\eea
Consequently, for example, 
\bea
e^{i\omega t(v)}\frac{dt}{dv} &=&e^{q\sinh v -p v}(e_r\cosh v-1)\bar a_r^{3/2}\times \nonumber\\
&\times& \left[ 1+(\tilde b_2+\tilde c_2)\frac{\eta^2}{\bar a_r}\right. \nonumber\\
&&\left.+(\tilde b_2 \tilde c_2+\tilde b_4+\tilde c_4)\frac{\eta^4}{\bar a_r^2}  \right]+O(\eta^6)\,.
\eea

At the Newtonian level the computation of the Fourier transform of the various multipole moments (see, e.g., Eq. \eqref{I_ab_omega2}) is done by using the integral representation of the Hankel functions of the first kind of order $p$ and argument $q$ (with Eqs. \eqref{defupq})
\beq
\label{Hankel_rep}
H_p^{(1)}(q)=\frac{1}{ i\pi }\int_{-\infty}^\infty e^{q\sinh v -p v}dv\,.
\eeq
As the argument $q=iu$  of the Hankel function is purely imaginary, the Hankel function reduces to a 
Bessel K function, according to the  relation
\beq
\label{H1vsK}
H_p^{(1)}(iu)=\frac{2}{\pi}e^{-i \frac{\pi}{2}(p+1)}K_p(u)\,.
\eeq
The typical term is of the kind $e^{q\sinh v -(p+k) v}$, the Fourier transform of which is
\beq
e^{q\sinh v -(p+k) v}\to2e^{-i \frac{\pi}{2}(p+k)}K_{p+k}(u)\,,
\eeq
involving Bessel functions having the same argument $u$, but various orders differing by integers.
However,  standard identities valid for Bessel functions allow one to reduce the orders to either $p$ or $p+1$. 

Higher orders in PN expansion imply for the integration in $v$ more complicated expressions. 
For instance, the Fourier transform of terms like $v^ne^{q\sinh v -(p+k) v}$ can be computed by
\beq
v^ne^{q\sinh v -(p+k) v}\to2(-1)^n\frac{\partial^n}{\partial p^n}\left[e^{-i \frac{\pi}{2}(p+k)}K_{p+k}(u)\right]\,.
\eeq
However, one generally has terms of the form $e^{q\sinh v -(p+k) v}f(v)$, which cannot be integrated analytically.
We will see below how to overcome this difficulty by integrating over the frequencies first.

\subsection{Newtonian result for the gravitational wave energy}

At the Newtonian order we find
\beq 
\label{EGWuN}
\Delta E_{\rm GW}^{\rm N}=\frac{G}{\pi c^5}\frac{1}{e_r \bar a_r^{3/2}}  \int_{0}^\infty  du  {\mathcal K}_{\rm N}(u)\,,
\eeq
with

\begin{widetext}

\bea
\label{calKNgen}
\mathcal K_{\rm N}(u)&=&
-\frac{32}{5}\frac{\nu^2}{\bar a_r^2}\frac{p^2}{u^4}e^{-i\pi p}\left\{
u^2(p^2+u^2+1)(p^2+u^2)K_{p+1}^2(u)\right.\nonumber\\
&&
-2u\left[\left(p-\frac32\right)u^2+p(p-1)^2\right](p^2+u^2)K_{p}(u)K_{p+1}(u)\nonumber\\
&&\left.
+2\left[\frac12u^6+\left(2p^2-\frac32p+\frac16\right)u^4+\left(\frac52p^4-\frac72p^3+p^2\right)u^2+p^4(p-1)^2\right]K_{p}^2(u)
\right\}\,.
\eea

\end{widetext}

The integral \eqref{EGWuN} cannot be performed analytically, since the order of the Bessel functions also depends on the integration variable.
This is the reason why no further step has been done for long time after the work \cite{Zeldovich:1974}, where only the asymptotic behavior was investigated in the limiting cases of large eccentricity ($e_r \to \infty$) and parabolic orbits ($e_r \to 1$) in both low-frequency and high-frequency regimes by using the asymptotic form of the Hankel functions, following a similar analysis done by Landau and Lifshitz in the case of electromagnetic radiation \cite{Landau:1982dva}.
The same computation has been done recently in Ref. \cite{DeVittori:2012da}, later corrected in Refs. \cite{Garcia-Bellido:2017knh,Grobner:2020fnb}.

However, the order $p=i u/e_r$ tends to zero when $e_r \to \infty$, so that one can take a large-eccentricity expansion of the integrand \eqref{calKNgen}, which allows for analytical computation beyond the leading order \cite{Bini:2017wfr}.
The notation LO, NLO, NNLO, ect. below refer to this large-eccentricity expansion.

\section{Large-eccentricity expansion}

When taking the large-eccentricity expansion one then expands with respect to the order of the Bessel functions.
This gives rise, at LO, to $K_0(u)$, and $K_1(u)$,  and at NLO, NNLO, N$^n$NLO, to derivatives of $K_0(u)$, and $K_1(u)$
with respect to their orders. 
Taylor-expanding $K_p(u)$ and $K_{p+1}(u)$ around $p=0$ gives
\bea
\label{K_exp}
K_p(u)&=&  K_0(u)+\frac12 p^2 \frac{\partial^2 K_\nu(u)}{\partial \nu^2}\Bigg|_{\nu=0}+O(p^3)
\,,\nonumber\\
K_{p+1}(u)&=& K_1(u)+\frac{p}{u} K_0(u) \nonumber\\
&&+\frac12 p^2 \frac{\partial^2 K_\nu(u)}{\partial \nu^2 }\Bigg|_{\nu=1}+O(p^3)\,,
\eea
to second order in $p$, where we have used the relations (see Eqs. 9.1.66-9.1-68 of Ref. \cite{AS})
\beq
\frac{\partial K_\nu(u)}{\partial \nu}\Bigg|_{\nu=0}=0\,,\qquad
\frac{\partial K_\nu(u)}{\partial \nu}\Bigg|_{\nu=1}=\frac{1}{u} K_0(u)\,.
\eeq

Each PN term entering the integrand \eqref{calKPNexp} can then be expanded as
\bea
\label{calKnPNexp}
\mathcal K_{\rm nPN}(u)&=&\frac{\nu^2}{e_r^2\bar a_r^2}\left[\tilde {\mathcal K}_{\rm nPN}^{\rm LO}(u)+\frac{\pi}{e_r}\tilde {\mathcal K}_{\rm nPN}^{\rm NLO}(u)\right.\nonumber\\
&+&\left.
\frac{1}{e_r^2}\tilde {\mathcal K}_{\rm nPN}^{\rm NNLO}(u)
+O\left(\frac{1}{e_r^3}\right)\right]\,,
\eea
up to the NNLO order in the large eccentricity.
We recall that the Fourier transform of the multipole moments cannot be always done in closed analytical form.
Therefore, starting at 1PN order the various terms entering the expansion \eqref{calKPNexp} may involve integrals over $v$ which cannot be explicitly evaluated.
However, it is still possible to analytically compute the resulting double integral over both $u$ and $v$ by integrating first over $u$ using the Mellin transform, and
then integrating over $v$.

\subsection{Integrating over the frequency spectrum and Mellin transform}
 
In order to compute the gravitational wave energy \eqref{EGWu} we have to evaluate the integral
\beq 
\label{Idef}
I_{\Delta E}=\int_{0}^\infty  du  {\mathcal K}(u)\,,
\eeq
where $\mathcal K(u)$ is given in the form of the double PN (i.e., $\eta$) + PM (i.e., $e_r^{-1}\sim j^{-1}\propto G$) expansion, Eqs. \eqref{calKPNexp} and \eqref{calKnPNexp}.
The various terms are listed in Appendix A.
The above integral then splits in different contributions, i.e.,
\beq
I_{\Delta E}=I_{\Delta E,\,\rm N}+\frac{\eta^2}{\bar a_r}I_{\Delta E,\,\rm 1PN}+\frac{\eta^4}{\bar a_r^2}I_{\Delta E,\,\rm 2PN}\,,
\eeq
further decomposing each term $I_{\Delta E,\,\rm nPN}$ as in Eq. \eqref{calKnPNexp}, which can be straightforwardly computed by using the Mellin transform of each function $\mathcal K_{\rm nPN}(u)$, as shown below.
We recall that the Mellin transform of a function $f(x)$ is defined as 
\beq
\label{mellin}
g(s)\equiv \mathfrak M\{f(x);s\}=\int_0^\infty x^{s-1} f(x) dx\,,
\eeq
with the property $\mathfrak M\{x^kf(x);s\}=g(s+k)$, so that for example
\beq
\int_0^\infty x^kf(x) dx=g(1+k)\,.
\eeq

At the Newtonian level, the function $\mathcal K_{\rm N}(u)$ is expressed in terms of modified Bessel functions of the second kind only, Eq. \eqref{calKNlist}. The typical term has the form 
\beq
\label{ukKK}
u^kK_\mu(u)K_\nu(u)\,,
\eeq
with $\mu,\nu=0,1$.
Therefore,  in order to evaluate the integral of \eqref{ukKK} it is enough to compute the Mellin transform $g_{K_\mu K_\nu}(s)$ of the function 
\beq
f_{K_\mu K_\nu}(u)=K_\mu(u)K_\nu(u)\,.
\eeq

At higher PN orders also appear terms like (see, e.g., Eq. \eqref{calI1PNlist})
\bea
&&a_k(v)u^kK_\nu(u)\cos(u\sinh v)\,,\nonumber\\
&&b_k(v)u^kK_\nu(u)\sin(u\sinh v)\,,
\eea
to be integrated over both $u$ and $v$.
Hence we also need the Mellin transforms $g_{K_\nu{\rm cos}}(s;v)$ and $g_{K_\nu{\rm sin}}(s;v)$ of the functions
\bea
f_{K_\nu{\rm cos}}(u,v)&=&K_\nu(u)\cos(u\sinh v)\,,\nonumber\\
f_{K_\nu{\rm sin}}(u,v)&=&K_\nu(u)\sin(u\sinh v)\,.
\eea

Mellin transforms are well implemented in standard symbolic algebra manipulators.
We find
\bea
\label{gdefs}
g_{K_\mu K_\nu}(s) &=& \frac{ 2^{s-3}}{\Gamma(s)}\nonumber\\
&\times& \Gamma\left(\frac{s+\mu+\nu}{2} \right)\Gamma\left(\frac{s-\mu+\nu}{2} \right)\nonumber\\
&\times &
\Gamma\left(\frac{s+\mu-\nu}{2} \right)\Gamma\left(\frac{s-\mu-\nu}{2} \right)
\,,\nonumber\\
g_{K_\nu{\rm cos}}(s;v) &=&  
\frac{2^{s-2}}{\cosh^{s-\nu}v}\Gamma\left(\frac{s+\nu}{2}\right) \Gamma\left(\frac{s-\nu}{2}\right)\nonumber\\
&\times&
{}_{2}F_{1}\left(\frac{1-s-\nu}{2}, \frac{s-\nu}{2};\frac12;\tanh^2v\right)
\,,\nonumber\\
g_{K_\nu{\rm sin}}(s;v)&=&
\frac{2^{s-1}\sinh v}{\cosh^{1+s+\nu}v}\nonumber\\
&\times&
\Gamma\left(\frac{s+\nu+1}{2}\right) \Gamma\left(\frac{s-\nu+1}{2}\right)\nonumber\\
&\times&
{}_{2}F_{1}\left(\frac{2-s+\nu}{2}, \frac{s+\nu+1}{2};\frac32;\tanh^2v\right)
\,, \nonumber\\
\eea
which we need for $s\geq2$ and $\mu,\nu=0,1$.
For each of them we also need the second derivative with respect to the order $\nu$ and higher derivatives for increasing PN accuracy as well as level of expansion in the eccentricity parameter (see, e.g., Eq. \eqref{calI2PNlist}), i.e., 
\bea
g_{K_\mu\partial^2K_\nu}(s)&=&\frac{\partial^2}{\partial \nu^2}g_{K_\mu K_\nu}(s)
\,,\nonumber\\
g_{\partial^2K_\nu{\rm cos}}(s;v)&=&\frac{\partial^2}{\partial \nu^2}g_{K_\nu{\rm cos}}(s;v)
\,,\nonumber\\
g_{\partial^2K_\nu{\rm sin}}(s;v)&=&\frac{\partial^2}{\partial \nu^2}g_{K_\nu{\rm sin}}(s;v)
\,,
\eea
which are the Mellin transforms of the functions
\bea
f_{K_\mu\partial^2K_\nu}(u)&=&K_\mu(u)\frac{\partial^2 K_\nu(u)}{\partial \nu^2}
\,,\nonumber\\
f_{\partial^2K_\nu{\rm cos}}(u,v)&=&\frac{\partial^2 K_\nu(u)}{\partial \nu^2}\cos(u\sinh v)
\,,\nonumber\\
f_{\partial^2K_\nu{\rm sin}}(u,v)&=&\frac{\partial^2 K_\nu(u)}{\partial \nu^2}\sin(u\sinh v)
\,,
\eea
respectively.

\subsection{Newtonian term $I_{\Delta E,\,\rm N}$}

The various contributions to the integral $I_{\Delta E,\,\rm N}$ are easily evaluated by replacing each term $u^kf_{K_\mu K_\nu}(u)$ and $u^kf_{K_\mu\partial^2K_\nu}(u)$ in Eq. \eqref{calKNlist} with the corresponding Mellin transforms $g_{K_\mu K_\nu}(1+k)$ and $g_{K_\mu\partial^2K_\nu}(1+k)$, respectively.
We find

\begin{widetext}

\bea
\label{IDeltaEN}
I_{\Delta E,\,\rm N}^{\rm LO}&=&\frac{32}{5}g_{K_0K_0}(5)+\frac{32}{15}g_{K_0K_0}(3)+\frac{96}{5}g_{K_0K_1}(4)+\frac{32}{5}g_{K_1K_1}(5)+\frac{32}{5}g_{K_1K_1}(3)
\nonumber\\
&=&\frac{37}{15}\pi^2 
\,,\nonumber\\
I_{\Delta E,\,\rm N}^{\rm NLO}&=&\frac{32}{5}g_{K_0K_0}(6)+\frac{32}{15}g_{K_0K_0}(4)+\frac{96}{5}g_{K_0K_1}(5)+\frac{32}{5}g_{K_1K_1}(6)+\frac{32}{5}g_{K_1K_1}(4)
\nonumber\\
&=&\frac{1568}{45}
\,,\nonumber\\
I_{\Delta E,\,\rm N}^{\rm NNLO}&=&\frac{16 \pi ^2}{5}g_{K_0K_0}(7)+\left(\frac{16 \pi ^2}{15}-\frac{96}{5}\right)g_{K_0K_0}(5)-\frac{32}{5}g_{K_0K_0}(3)
+\frac{48 \pi ^2}{5}g_{K_0K_1}(6)-\frac{224 }{5}g_{K_0K_1}(4)\nonumber\\
&&
+\frac{16 \pi ^2}{5}g_{K_1K_1}(7)+\left(\frac{16 \pi ^2}{5}-\frac{64}{5}\right) g_{K_1K_1}(5)-\frac{32}{5}g_{K_1K_1}(3)
-\frac{32 }{5}g_{K_0\partial^2K_0}(7)-\frac{32 }{15}g_{K_0\partial^2K_0}(5)\nonumber\\
&&
-\frac{48 }{5}(g_{K_1\partial^2K_0}(6)+g_{K_0\partial^2K_1}(6))-\frac{32}{5}g_{K_1\partial^2K_1}(7)-\frac{32}{5}g_{K_1\partial^2K_1}(5)
\nonumber\\
&=&\frac{281}{10}\pi^2
\,,
\eea
where we have used
\bea
g_{K_0K_0}(7)&=&\frac{1125\pi^2}{4096}\,,\qquad
g_{K_0K_0}(6)=\frac{16}{15}\,,\qquad
g_{K_0K_0}(5)=\frac{27\pi^2}{512}\,,\qquad
g_{K_0K_0}(4)=\frac{1}{3}\,,\qquad
g_{K_0K_0}(3)=\frac{\pi^2}{32}
\,,\nonumber\\
g_{K_0K_1}(6)&=&\frac{135\pi^2}{1024}\,,\qquad
g_{K_0K_1}(5)=\frac{2}{3}\,,\qquad
g_{K_0K_1}(4)=\frac{3\pi^2}{64}
\,,\nonumber\\
g_{K_1K_1}(7)&=&\frac{1575\pi^2}{4096}\,,\qquad
g_{K_1K_1}(6)=\frac{8}{5}\,,\qquad
g_{K_1K_1}(5)=\frac{45\pi^2}{512}\,,\qquad
g_{K_1K_1}(4)=\frac{2}{3}\,,\qquad
g_{K_1K_1}(3)=\frac{3\pi^2}{32}
\,,\nonumber\\
g_{K_0\partial^2K_0}(7)&=&\frac{5\pi^2(-2072+225\pi^2)}{8192} \,,\qquad
g_{K_0\partial^2K_0}(5)=\frac{3\pi^2 (-80+9\pi^2)}{1024}
\,,\nonumber\\
g_{K_1\partial^2K_0}(6)&=&\frac{3\pi^2(-2036+225\pi^2)}{10240}\,,\qquad
g_{K_0\partial^2K_1}(6)=\frac{3\pi^2(-1964+225\pi^2) }{10240}
\,,\nonumber\\
g_{K_1\partial^2K_1}(7)&=&\frac{\pi^2(-14072+1575\pi^2)}{8192}\,,\qquad
g_{K_1\partial^2K_1}(5)=\frac{\pi^2(-368+45\pi^2)}{1024}
\,.
\eea

\end{widetext}

Therefore, the Newtonian value of the gravitational wave energy \eqref{EGWuN} up to the NNLO in the large-eccentricity expansion reads 
\bea 
\label{deltaENfin}
\Delta E_{\rm GW}^{\rm N}&=&\frac{G}{ c^5}\frac{\nu^2}{e_r^3 \bar a_r^{7/2}}\left[
\frac{37}{15}\pi + \frac{1}{e_r}\frac{1568}{45} + \frac{1}{e_r^2}\frac{281}{10}\pi\right.\nonumber\\
&&\left.
+O\left(\frac{1}{e_r^3}\right)\right]\,.
\eea

\subsection{1PN term $I_{\Delta E,\,\rm 1PN}$}

Let us pass to the 1PN term $I_{\Delta E,\,\rm 1PN}$.
Most of the contributions in Eq. \eqref{calK1PNlist} can be computed exactly as in the Newtonian case, leading to
\bea
\label{IDeltaE1PN}
I_{\Delta E,\,\rm 1PN}^{\rm LO}&=&\left(\frac{1143}{280}-\frac{37}{30}\nu\right)\pi^2
\,,\nonumber\\
I_{\Delta E,\,\rm 1PN}^{\rm NLO}&=&-\frac{25616}{315}-\frac{1136}{45}\nu
+\int_0^{\infty} du{\mathcal I}_{\rm 1PN}^{\rm NLO}(u)
\,,\nonumber\\
I_{\Delta E,\,\rm 1PN}^{\rm NNLO}&=&\left(-\frac{111037}{560}-\frac{609}{20}\nu\right)\pi^2
+\int_0^{\infty} du{\mathcal I}_{\rm 1PN}^{\rm NNLO}(u)
\,,\nonumber\\
\eea
where two integrals still remain to be evaluated.
Consider the NLO integral in the above equation.
Integrating over $u$ first, i.e., substituting each term $u^kf_{K_\nu{\rm cos}}(u)$ and $u^kf_{K_\nu{\rm sin}}(u)$ with the corresponding Mellin transforms $g_{K_\nu{\rm cos}}(1+k)$ and $g_{K_\nu{\rm sin}}(1+k)$ then yields

\begin{widetext}

\bea
\int_0^{\infty} du{\mathcal I}_{\rm 1PN}^{\rm NLO}(u)&=&-\frac{96}{5\pi}\int_{-\infty}^{\infty} dv\,{\rm arctan}\left(\tanh\frac{v}{2}\right)\cosh v\left[
\sinh v(g_{K_0{\rm cos}}(5;v)+2 g_{K_1{\rm cos}}(6;v))\right.\nonumber\\
&&\left.
+ (\cosh^2v-2)(g_{K_0{\rm sin}}(6;v)+g_{K_1{\rm sin}}(5;v))
\right]\nonumber\\
&=&\int dv\, {\rm arctan}\left(\tanh\left(\frac{v}{2}\right)\right)\frac{\sinh v}{\cosh^4 v}\left(-\frac{4032}{5}+\frac{2448}{\cosh^2 v}\right)\nonumber\\
&=&\frac{2048}{25}\,,
\eea
where we have used 
\bea
g_{K_0{\rm cos}}(5;v)&=&\frac{3\pi}{2\cosh^9v} (8 \cosh^4v-40 \cosh^2v+35)
\,,\nonumber\\
g_{K_1{\rm cos}}(6;v)&=&\frac{45\pi}{2\cosh^{11}v} (8 \cosh^4v-28 \cosh^2v+21)
\,,\nonumber\\
g_{K_0{\rm sin}}(6;v)&=&\frac{15\pi\sinh v}{2\cosh^{11}v}  (8 \cosh^4v-56 \cosh^2v+63) 
\,,\nonumber\\
g_{K_1{\rm sin}}(5;v)&=&-\frac{15\pi\sinh v}{2\cosh^9v} (4 \cosh^2v-7) 
\,,
\eea
so that 
\beq
I_{\Delta E,\,\rm 1PN}^{\rm NLO}=\frac{944}{1575}-\frac{1136}{45}\nu\,.
\eeq

In the same way the remaining NNLO integral in Eq. \eqref{IDeltaE1PN} turns out to be
\beq
\int_0^{\infty} du{\mathcal I}_{\rm 1PN}^{\rm NNLO}(u)=\frac{792}{5}\pi^2\,,
\eeq
so that 
\beq
I_{\Delta E,\,\rm 1PN}^{\rm NNLO}=\left(-\frac{22333}{560}-\frac{609}{20}\nu\right)\pi^2\,.
\eeq

Summarizing, the 1PN value of the gravitational wave energy up to the NNLO in the large-eccentricity expansion reads 
\beq 
\label{deltaE1PNfin}
\Delta E_{\rm GW}^{\rm 1PN}=\frac{G}{ c^5}\frac{\nu^2}{e_r^3 \bar a_r^{9/2}}\left[
\left(\frac{1143}{280}-\frac{37}{30}\nu\right)\pi 
+ \frac{1}{e_r}\left(\frac{944}{1575}-\frac{1136}{45}\nu\right) 
+ \frac{1}{e_r^2}\left(-\frac{22333}{560}-\frac{609}{20}\nu\right)\pi
+O\left(\frac{1}{e_r^3}\right)\right]\,.
\eeq

\subsection{2PN term $I_{\Delta E,\,\rm 2PN}$}

The computation of the 2PN term $I_{\Delta E,\,\rm 2PN}$ proceeds exactly as before.
We find
\bea
\label{IDeltaE2PN}
I_{\Delta E,\,\rm 2PN}^{\rm LO}&=&\left(\frac{158539}{20160}-\frac{7443}{560}\nu+\frac{1607}{560}\nu^2\right)\pi^2\nonumber\\
&&
+\int_0^{\infty} du{\mathcal I}_{\rm 2PN}^{\rm LO}(u)
\,,\nonumber\\
I_{\Delta E,\,\rm 2PN}^{\rm NLO}&=&-\frac{178768}{1575}+\frac{311264}{1575}\nu+\frac{13504}{1575}\nu^2\nonumber\\
&&
+\int_0^{\infty} du{\mathcal I}_{\rm 2PN}^{\rm NLO}(u)
\,,\nonumber\\
I_{\Delta E,\,\rm 2PN}^{\rm NNLO}&=&\left[
\left(\frac{297 \pi ^2}{640}-\frac{820373}{16800}\right) \nu ^2+\left(\frac{3113597}{3360}-\frac{2673 \pi ^2}{320}\right) \nu +\frac{24057 \pi ^2}{640}-\frac{98759779}{201600}
\right]\pi^2\nonumber\\
&&
+\int_0^{\infty} du{\mathcal I}_{\rm 2PN}^{\rm NNLO}(u)
\,,\nonumber\\
\eea
where the quantities ${\mathcal I}_{\rm 2PN}^{\rm LO}(u)$, ${\mathcal I}_{\rm 2PN}^{\rm NLO}(u)$ and ${\mathcal I}_{\rm 2PN}^{\rm NNLO}(u)$ are listed in Eq. \eqref{calI2PNlist}.
Integrating the remaining LO integral over $u$ and then over $v$ gives
\bea
\int_0^{\infty} du{\mathcal I}_{\rm 2PN}^{\rm LO}(u)&=&\int_{-\infty}^{\infty} dv\,\frac{\pi}{\cosh^3v}\left\{
v\tanh v\left[\frac{-9270 \nu ^2+76230 \nu +64800}{\cosh^6v}
+\frac{10191 \nu ^2-87563 \nu -75519}{\cosh^4v}\right.\right.\nonumber\\
&&\left.
+\frac{-\frac{14464 \nu ^2}{7}+18816 \nu+\frac{115536}{7}}{\cosh^2v}
-\frac{632 \nu ^2}{7}+872 \nu +\frac{5496}{7}\right]\nonumber\\
&&
+\frac{64920 \nu ^2+110220 \nu +41730}{\cosh^8v}
+\frac{-\frac{7634194 \nu ^2}{63}-\frac{15656588 \nu }{63}-\frac{6364793}{63}}{\cosh^6v}\nonumber\\
&&
+\frac{\frac{6705371 \nu ^2}{105}+\frac{17568433\nu }{105}+\frac{7846411}{105}}{\cosh^4v}
+\frac{-\frac{232153 \nu ^2}{35}-\frac{2850973 \nu }{105}-\frac{1511878}{105}}{\cosh^2v}\nonumber\\
&&\left.
-\frac{199226 \nu ^2}{315}-\frac{774814 \nu}{315}-\frac{352816}{315}
\right\}\nonumber\\
&=&\left(\frac{23221}{6720}+\frac{1091}{120}\nu-\frac{1089}{560}\nu^2\right)\pi^2
\,,
\eea
so that 
\beq
I_{\Delta E,\,\rm 2PN}^{\rm LO}=\left(\frac{114101}{10080}-\frac{1411}{336}\nu+\frac{37}{40}\nu^2\right)\pi^2
\,.
\eeq

Similarly, the NLO integral turns out to be
\beq
\int_0^{\infty} du{\mathcal I}_{\rm 2PN}^{\rm NLO}(u)=\frac{556804}{3675}-\frac{359924}{1575}\nu+\frac{4412}{525}\nu^2
\,,
\eeq
so that
\beq
I_{\Delta E,\,\rm 2PN}^{\rm NLO}=\frac{419036}{11025}-\frac{3244}{105}\nu+\frac{764}{45}\nu^2
\,.
\eeq

Finally, the NNLO integral turns out to be
\bea
\label{intucalI2PNNNLO}
\int_0^{\infty} du{\mathcal I}_{\rm 2PN}^{\rm NNLO}(u)&=&\left[
\left(\frac{1135163}{16800}-\frac{297 \pi ^2}{640}\right) \nu ^2+\left(\frac{2673 \pi ^2}{320}-\frac{728113}{840}\right) \nu -\frac{24057 \pi ^2}{640}+\frac{26607403}{67200}
\right]\pi^2\nonumber\\
&&
+\frac{72}{5}\int_0^{\infty} du\,u^6\int_{-\infty}^{\infty} dv\int_{-\infty}^{\infty} dv'\, {\rm arctan}\left(\tanh\frac{v}{2}\right){\rm arctan}\left(\tanh\frac{v'}{2}\right)
e^{iu(\sinh v-\sinh v')}\nonumber\\
&&\times
\cosh v\cosh v'[\cosh^2v\cosh^2v'-2(\sinh v -\sinh v')^2]
\,.
\eea
In order to compute the latter integral it is convenient to change variables as $\sinh v=x$ and $\sinh v'=x'$, leading to 
\beq
\frac{72}{5}\int_0^{\infty} du\,u^6\int_{-\infty}^{\infty} dx\int_{-\infty}^{\infty} dx'\, {\rm arctan}\left(\frac{\sqrt{1+x^2}-1}{x}\right){\rm arctan}\left(\frac{\sqrt{1+x'^2}-1}{x'}\right)e^{iu(x-x')}[(1+xx')^2-(x-x')^2]\,,
\eeq
integrating then over $u$ by using the integral representation of the Dirac delta function $\int_{-\infty}^{\infty} dp\, e^{ip(x-x_0)}=2\pi \delta(x-x_0)$, which yields
\bea
&&
-\frac{72}{5}\pi\int_{-\infty}^{\infty} dx\int_{-\infty}^{\infty} dx'\, \frac{\partial^6}{\partial x^6}\left[{\rm arctan}\left(\frac{\sqrt{1+x^2}-1}{x}\right){\rm arctan}\left(\frac{\sqrt{1+x'^2}-1}{x'}\right)[(1+xx')^2-(x-x')^2]\right]\delta(x-x')\nonumber\\
&&
=\frac{6912}{5}\pi\int_{-\infty}^{\infty} dv\, {\rm arctan}\left(\tanh\frac{v}{2}\right)\frac{\sinh v}{\cosh^7v}(11-5\cosh^2v)\nonumber\\
&&
=72\pi^2
\,.
\eea
Therefore, the NNLO integral \eqref{intucalI2PNNNLO} finally reads
\beq
\int_0^{\infty} du{\mathcal I}_{\rm 2PN}^{\rm NNLO}(u)=\left[
\left(\frac{1135163}{16800}-\frac{297 \pi ^2}{640}\right) \nu ^2+\left(\frac{2673 \pi ^2}{320}-\frac{728113}{840}\right) \nu -\frac{24057 \pi ^2}{640}+\frac{31445803}{67200}
\right]\pi^2\,,
\eeq
so that
\beq
I_{\Delta E,\,\rm 2PN}^{\rm NNLO}=\left(-\frac{442237}{20160}+\frac{5747}{96}\nu+\frac{1499}{80}\nu^2\right)\pi^2
\,.
\eeq

Summarizing, the 2PN value of the gravitational wave energy up to the NNLO in the large-eccentricity expansion reads 
\bea 
\label{deltaE2PNfin}
\Delta E_{\rm GW}^{\rm 2PN}&=&\frac{G}{ c^5}\frac{\nu^2}{e_r^3 \bar a_r^{11/2}}\left[
\left(\frac{114101}{10080}-\frac{1411}{336} \nu +\frac{37}{40} \nu ^2\right)\pi 
+ \frac{1}{e_r}\left(\frac{419036}{11025}-\frac{3244}{105} \nu +\frac{764 }{45}\nu ^2\right)\right.\nonumber\\
&&\left.
+ \frac{1}{e_r^2}\left(-\frac{442237}{20160}+\frac{5747}{96} \nu +\frac{1499 }{80}\nu ^2\right)\pi
+O\left(\frac{1}{e_r^3}\right)\right]\,.
\eea

\subsection{Large-$j$ expansion}

It is useful to express the above result as a series in inverse powers of the large angular momentum.
Collecting all terms together and replacing $(\bar a_r,e_r)$ by $(\bar E,j)$ through Eq. \eqref{arer} finally gives
\bea 
\label{deltaE2PNlargej}
\Delta E_{\rm GW}&=&\frac{G}{ c^5}\frac{\nu^2p_\infty^4}{j^3}\left\{
\left[\frac{37}{15}+\left(\frac{1357}{840}-\frac{74}{15}\nu\right)p_\infty^2+\left(\frac{27953}{10080}-\frac{839}{420}\nu+\frac{37}{5}\nu^2\right)p_\infty^4
+O(p_\infty^6)\right]\pi\right.\nonumber\\
&&
+\left[\frac{1568}{45}+\left(\frac{18608}{525}-\frac{1424}{15}\nu\right)p_\infty^2+\left(\frac{220348}{11025}-\frac{31036}{525}\nu+172\nu^2\right)p_\infty^4
+O(p_\infty^6)\right]\frac{1}{p_\infty j}\nonumber\\
&&
+\left[\frac{122}{5}+\left(\frac{13831}{280}-\frac{933}{10}\nu\right)p_\infty^2+\left(-\frac{64579}{5040}-\frac{187559}{1680}\nu+\frac{2067}{10}\nu^2\right)p_\infty^4+O(p_\infty^6)\right]\frac{\pi}{(p_\infty j)^2}\nonumber\\
&&\left.
+O\left(\frac{1}{(p_\infty j)^3}\right)
\right\}\,,
\eea
having used the relation between binding energy $\bar E$ and momentum at infinity $p_\infty$,  as given by
\beq
\bar E= \frac{E_{\rm tot}-Mc^2}{\mu c^2}=\frac{h(\gamma,\nu)-1}{\nu}\,,\qquad h(\gamma,\nu)=\sqrt{1+2\nu(\gamma-1)}\,,\qquad \gamma=\sqrt{1+p_\infty^2\eta^2}\,,
\eeq
that is
\beq
\bar E=\frac12p_\infty^2-\frac18(1+\nu)p_\infty^4\eta^2+\frac{1}{16}(1+\nu+\nu^2)p_\infty^6\eta^4+O(\eta^6)\,.
\eeq

\subsection{Exact 2PN result in the time domain}

We recall below the current knowledge of the total gravitational-wave energy emitted during a scattering process, recently computed in a previous work \cite{Bini:2020hmy} in the time domain to the 2PN accuracy, thereby generalizing the 1PN-accurate result of Ref. \cite{Blanchet:1989cu}.
When expressed in terms of $(\bar a_r, e_r)$ instead of $(e_r,j)$ through the relation 
\beq
\bar a_r(e_r^2-1)=j^2+[-4+(\nu-2)e_r^2]\eta^2
+\frac1{j^2}\left[9\nu-9+\left(-3+\frac{61}{4}\nu+\nu^2\right)e_r^2+\left(-4+\frac{15}{4}\nu-\nu^2\right)e_r^4\right]\eta^4\,,
\eeq
it reads
\beq
\label{deltaE2PNexact}
\Delta E_{\rm GW}= \frac{G}{ c^5}\frac{\nu^2}{\bar a_r^{7/2}(e_r^2-1)^{7/2}}\left({\mathcal E}^{\rm N}+\frac{\eta^2}{\bar a_r(e_r^2-1)}{\mathcal E}^{\rm 1PN}+\frac{\eta^4}{\bar a_r^2(e_r^2-1)^2}{\mathcal E}^{\rm 2PN}\right)\,,
\eeq
where
\beq
{\mathcal E}^{\rm nPN}= A^{\rm nPN}\,{\rm arccos}\left(-\frac{1}{e_r}\right)+B^{\rm nPN}\sqrt{e_r^2-1}\,,
\eeq
with coefficients
\bea
\label{deltaE2PNexactcoeffs}
A^{\rm N}&=&\frac{64}{5}+\frac{584}{15}e_r^2+\frac{74}{15}e_r^4 
\,,\nonumber\\
B^{\rm N}&=&\frac{1204}{45}+\frac{1346}{45}e_r^2 
\,,\nonumber\\
A^{\rm 1PN}&=&  -\frac{112}{5}\nu-\frac{1132}{21}
+\left(-\frac{37318}{105}-\frac{308}{3}\nu\right) e_r^2
+\left(-\frac{233}{2}-\frac{249}{5}\nu\right) e_r^4
+\left(\frac{1143}{140}-\frac{37}{15}\nu\right) e_r^6 
\,,\nonumber\\
B^{\rm 1PN}&=& -\frac{446}{9}\nu-\frac{210811}{1575}
+\left(-105\nu-\frac{592573}{1575}\right) e_r^2
+\left(-\frac{205}{9}\nu-\frac{47659}{6300}\right) e_r^4 
\,,\nonumber\\
A^{\rm 2PN}&=& \frac{105146}{315}\nu+\frac{32}{5}\nu^2+\frac{44134}{405}
+\left(\frac{387220}{567}+\frac{1089412}{315}\nu+\frac{623}{15}\nu^2\right) e_r^2\nonumber\\
&&
+\left(-\frac{108326}{945}+\frac{622831}{210}\nu+\frac{587}{12}\nu^2\right) e_r^4
+\left(-\frac{424337}{2520}+\frac{273}{10}\nu^2+\frac{27875}{168}\nu\right) e_r^6\nonumber\\
&&
+\left(\frac{37}{20}\nu^2+\frac{114101}{5040}-\frac{1411}{168}\nu\right) e_r^8 
\,,\nonumber\\
B^{\rm 2PN}&=& \frac{607888}{675}\nu+\frac{151}{10}\nu^2+\frac{78464696}{297675}
+\left(\frac{9047}{180}\nu^2+\frac{84265357}{18900}\nu+\frac{164159833}{238140}\right) e_r^2\nonumber\\
&&
+\left(\frac{59688863}{37800}\nu-\frac{520075147}{1190700}+\frac{2048}{45}\nu^2\right) e_r^4
+\left(-\frac{6299}{280}\nu+\frac{2711041}{176400}+\frac{2723}{180}\nu^2\right) e_r^6
\,.
\eea

\end{widetext}

It is easy to check that the large eccentricity expansion of the above expression reproduces the terms \eqref{deltaENfin}, \eqref{deltaE1PNfin}, and \eqref{deltaE2PNfin} computed above in the frequency domain. 

The method developed above for the energy released via gravitational waves in a gravitational two-body scattering problem can be used in any similar context. In Appendix \ref{LL} we will derive the dipolar energy released via electromagnetic waves in an electromagnetic two-body scattering process in the Fourier domain.
This problem is discussed in the classical textbook of Landau and Lifshitz \cite{Landau:1982dva} (see Chap. 9, Par. 70).
The Fourier computation of the energy radiated during the collision involves Hankel functions also in this case.
Therefore, only the limiting behavior of the spectrum for both low and high frequencies are studied, by using suitable approximations for the Hankel functions.
We will also show  how to get the total dipole radiation emitted during the whole scattering process by performing the integration over all frequencies as a power series in a large-eccentricity expansion.

\section{Concluding remarks}

We have computed in the frequency domain the averaged gravitational wave energy emitted during hyperbolic encounters in a two-body system up to the 2PN order and to the NNLO in a large eccentricity expansion.
Previous studies were limited to the Newtonian level and for large values of the impact parameter, corresponding to the leading order in the large eccentricity. 
In fact, the presence in the Fourier integrals of quadratic products of modified Bessel functions of the first kind with frequency-dependent order does not allow for computation in closed analytical form.
This is the reason why no further step has been done for long time.

We have shown how to evaluate these integrals by expanding the Bessel functions in power series of the large eccentricity, corresponding to small values of the order, so that the power spectrum eventually contains only Bessel functions of integer order at the Newtonian level.
Going beyond, the derivatives of the Bessel functions with respect to the order also appear, up to the second at the NNLO and higher derivatives for increasing PN accuracy as well as level of expansion in the eccentricity parameter.
A technical difficulty arising at 1PN order is the fact that the Fourier transform of the multipole moments from which the gravitational wave energy flux is constructed cannot be always done in closed form.
However, we have shown that it is still possible to analytically compute the resulting double integral by integrating over the frequencies first using the Mellin transform.

The procedure outlined here can be used then to reach higher N$^n$LO terms, as well as to compute also tail contributions to the energy flux, which we leave for future work.

\section*{Acknowledgments}
The authors thank T. Damour for useful discussions.
DB thanks the International Center for Relativistic Astrophysics Network, ICRANet, for partial support.
DB also acknowledges the hospitality  and the highly stimulating environment  of the Institut des Hautes Etudes Scientifiques. 
DB and AG thank MaplesoftTM for providing a complimentary license of Maple 2020.

\appendix

\section{Large-eccentricity expansion of $\Delta E_{\rm GW}$}

The total gravitational wave energy  emitted by the entire hyperbolic motion in the frequency domain reads (see Eqs. \eqref{EGWu}--\eqref{calKPNexp}) 
\beq 
\Delta E_{\rm GW}=\frac{G}{\pi c^5}\frac{1}{e_r \bar a_r^{3/2}}  \int_{0}^\infty  du  {\mathcal K}(u)\,,
\eeq
with
\beq
\mathcal K(u)=\mathcal K_{\rm N}(u)+\frac{\eta^2}{\bar a_r}\mathcal K_{\rm 1PN}(u)+\frac{\eta^4}{\bar a_r^2}\mathcal K_{\rm 2PN}(u)\,.
\eeq
Each term has the following large-eccentricity expansion (see Eq. \eqref{calKnPNexp})
\bea
\mathcal K_{\rm nPN}(u)&=&\frac{\nu^2}{e_r^2\bar a_r^2}\left[\tilde {\mathcal K}_{\rm nPN}^{\rm LO}(u)+\frac{\pi}{e_r}\tilde {\mathcal K}_{\rm nPN}^{\rm NLO}(u)\right.\nonumber\\
&+&\left.
\frac{1}{e_r^2}\tilde {\mathcal K}_{\rm nPN}^{\rm NNLO}(u)
+O\left(\frac{1}{e_r^3}\right)\right]\,,
\eea
up to the NNLO.
The various contributions are listed below.

We recall the definitions of the functions
\bea
f_{K_\mu K_\nu}(u)&=&K_\mu(u)K_\nu(u)
\,,\nonumber\\
f_{K_\nu{\rm cos}}(u,v)&=&K_\nu(u)\cos(u\sinh v)
\,,\nonumber\\
f_{K_\nu{\rm sin}}(u,v)&=&K_\nu(u)\sin(u\sinh v)\,,
\eea
with $\mu,\nu=0,1$, and their second derivatives with respect to the order $\nu$, i.e.,
\bea
f_{K_\mu\partial^2K_\nu}(u)&=&K_\mu(u)\frac{\partial^2 K_\nu(u)}{\partial \nu^2}
\,,\nonumber\\
f_{\partial^2K_\nu{\rm cos}}(u,v)&=&\frac{\partial^2 K_\nu(u)}{\partial \nu^2}\cos(u\sinh v)
\,,\nonumber\\
f_{\partial^2K_\nu{\rm sin}}(u,v)&=&\frac{\partial^2 K_\nu(u)}{\partial \nu^2}\sin(u\sinh v)
\,.
\eea

At the Newtonian level we find

\begin{widetext}

\bea
\label{calKNlist}
\tilde {\mathcal K}_{\rm N}^{\rm LO}(u)&=&\left(\frac{32 u^4}{5}+\frac{32 u^2}{15}\right)f_{K_0K_0}(u)+\frac{96}{5}u^3f_{K_0K_1}(u)+\left(\frac{32 u^4}{5}+\frac{32 u^2}{5}\right)f_{K_1K_1}(u)
\,,\nonumber\\
\tilde {\mathcal K}_{\rm N}^{\rm NLO}(u)&=&\left(\frac{32 u^5}{5}+\frac{32 u^3}{15}\right)f_{K_0K_0}(u)+\frac{96}{5}u^4f_{K_0K_0}(u)+\left(\frac{32 u^5}{5}+\frac{32 u^3}{5}\right)f_{K_0K_0}(u)
\,,\nonumber\\
\tilde {\mathcal K}_{\rm N}^{\rm NNLO}(u)&=&\left[\frac{16 \pi ^2 u^6}{5}+\left(\frac{16 \pi ^2}{15}-\frac{96}{5}\right) u^4-\frac{32u^2}{5}\right]f_{K_0K_0}(u)
+\left(\frac{48 \pi ^2 u^5}{5}-\frac{224 u^3}{5}\right)f_{K_0K_1}(u)\nonumber\\
&&
+\left[\frac{16 \pi ^2 u^6}{5}+\left(\frac{16 \pi ^2}{5}-\frac{64}{5}\right) u^4-\frac{32 u^2}{5}\right]f_{K_1K_1}(u)
+\left(-\frac{32 u^6}{5}-\frac{32 u^4}{15}\right)f_{K_0\partial^2K_0}(u)\nonumber\\
&&
-\frac{48 u^5}{5}(f_{K_1\partial^2K_0}(u)+f_{K_0\partial^2K_1}(u))+\left(-\frac{32 u^6}{5}-\frac{32 u^4}{5}\right)f_{K_1\partial^2K_1}(u)
\,.
\eea

At the 1PN level we find
\bea
\label{calK1PNlist}
\tilde{\mathcal K}_{\rm 1PN}^{\rm LO}(u)&=&\left[\left(\frac{16}{21}-\frac{64 \nu }{21}\right) u^6-\frac{656 u^4}{105}+\left(\frac{128 \nu }{105}-\frac{416}{35}\right)u^2\right]f_{K_0K_0}(u)\nonumber\\
&&
	+\left[\left(\frac{128 \nu }{35}+\frac{1952}{105}\right)u^5+\left(\frac{512 \nu }{105}-\frac{1376}{105}\right) u^3\right]f_{K_0K_1}(u)\nonumber\\
&&
	+\left[\left(\frac{16}{21}-\frac{64\nu }{21}\right) u^6+\left(\frac{32 \nu }{105}+\frac{24}{7}\right) u^4+\frac{160 u^2}{7}\right]f_{K_1K_1}(u)
\,,\nonumber\\
\tilde{\mathcal K}_{\rm 1PN}^{\rm NLO}(u)&=&\left[\left(\frac{16}{21}-\frac{64 \nu }{21}\right) u^7+\left(\frac{16 \nu }{5}-\frac{736}{21}\right)u^5+\left(\frac{16 \nu }{7}-\frac{752}{35}\right) u^3\right]f_{K_0K_0}(u)\nonumber\\
&&
+\left[\left(\frac{128 \nu}{35}+\frac{1952}{105}\right) u^6+\left(\frac{304 \nu }{21}-\frac{10448}{105}\right) u^4\right]f_{K_0K_1}(u)\nonumber\\
&&
+\left[\left(\frac{16}{21}-\frac{64 \nu }{21}\right) u^7+\left(\frac{368 \nu }{105}-\frac{888}{35}\right)u^5+\left(\frac{16 \nu }{5}-\frac{208}{35}\right) u^3\right]f_{K_1K_1}(u)\nonumber\\
&&
+{\mathcal I}_{\rm 1PN}^{\rm NLO}(u)
\,,\nonumber\\
\tilde{\mathcal K}_{\rm 1PN}^{\rm NNLO}(u)&=&\left\{\pi ^2 \left(\frac{8}{21}-\frac{32 \nu }{21}\right) u^8
+\left[-\frac{80}{21}+\frac{320 \nu }{21}+\pi ^2 \left(\frac{16 \nu }{5}-\frac{3352}{105}\right)\right]u^6\right.\nonumber\\
&&\left.
+\left[\frac{8728}{105}-\frac{1088 \nu }{105}+\pi ^2 \left(\frac{176 \nu }{105}-\frac{544}{35}\right)\right] u^4
+\left(\frac{1216}{35}-\frac{32 \nu }{5}\right)u^2\right\}f_{K_0K_0}(u)\nonumber\\
&&
+\left\{\pi ^2 \left(\frac{64 \nu }{35}+\frac{976}{105}\right) u^7
+\left[-\frac{2272}{35}+\frac{128 \nu }{5}+\pi ^2 \left(\frac{1264 \nu}{105}-\frac{1952}{21}\right)\right] u^5
+\left(\frac{3968}{35}-\frac{160 \nu }{7}\right) u^3\right\}f_{K_0K_1}(u)\nonumber\\
&&
+\left\{\pi ^2 \left(\frac{8}{21}-\frac{32 \nu}{21}\right) u^8
+\left[-\frac{64}{21}+\frac{256 \nu }{21}+\pi ^2 \left(\frac{352 \nu }{105}-\frac{948}{35}\right)\right] u^6\right.\nonumber\\
&&\left.
+\left[\frac{608}{105}+\frac{384 \nu }{35}+\pi ^2 \left(\frac{16\nu }{5}-\frac{608}{35}\right)\right] u^4
+\left(\frac{32 \nu }{7}-\frac{128}{5}\right) u^2\right\}f_{K_1K_1}(u)\nonumber\\
&&
+\left[\left(\frac{64 \nu }{21}-\frac{16}{21}\right) u^8+\left(\frac{6704}{105}-\frac{32 \nu }{5}\right) u^6+\left(\frac{1088}{35}-\frac{352 \nu }{105}\right)
   u^4\right]f_{K_0\partial^2K_0}(u)\nonumber\\
&&
+\left[\left(-\frac{64 \nu }{35}-\frac{976}{105}\right) u^7+\left(\frac{1952}{21}-\frac{1264 \nu }{105}\right) u^5\right](f_{K_1\partial^2K_0}(u)+f_{K_0\partial^2K_1}(u))\nonumber\\
&&
+\left[\left(\frac{64 \nu }{21}-\frac{16}{21}\right) u^8+\left(\frac{1896}{35}-\frac{704 \nu }{105}\right) u^6+\left(\frac{1216}{35}-\frac{32 \nu }{5}\right) u^4\right]f_{K_1\partial^2K_1}(u)\nonumber\\
&&
+{\mathcal I}_{\rm 1PN}^{\rm NNLO}(u)
\,,
\eea
with
\bea
\label{calI1PNlist}
{\mathcal I}_{\rm 1PN}^{\rm NLO}(u)&=&-\frac{96}{5\pi}\int_{-\infty}^{\infty} dv\left[
A\sh\ch (u^4 f_{K_0{\rm cos}}(u,v)+2 u^5 f_{K_1{\rm cos}}(u,v))
+ A\ch(\ch^2-2)(u^5 f_{K_0{\rm sin}}(u,v)+u^4 f_{K_1{\rm sin}}(u,v))
\right]
\,,\nonumber\\
{\mathcal I}_{\rm 1PN}^{\rm NNLO}(u)&=&
-\frac{48}{5}\int_{-\infty}^{\infty} dv\left\{
\left[-2 A C v \left(C^2-2\right) u^6+\pi  A C S u^5+S \left(S-2 A \left(C^2+1\right)\right) u^4\right]f_{K_0{\rm cos}}(u,v)\right.\nonumber\\
&&
+\left[2 \pi  A C S u^6+\left(A \left(-4 \left(C^2+1\right) S-2 C \left(C^2-2\right) v\right)+2 S^2\right)u^5 \right]f_{K_1{\rm cos}}(u,v)
\nonumber\\
&&
+\left[\pi  A C \left(C^2-2\right) u^6+\left(A \left(2 C^2+2 C S v+4\right)+\left(C^2-2\right) S\right)u^5 \right]f_{K_0{\rm sin}}(u,v)\nonumber\\
&&\left.
+\left[4 A C S v u^6  +\pi  A C \left(C^2-2\right) u^5+ \left(A \left(2 C^2+4\right)+\left(C^2-2\right) S\right)u^4\right]f_{K_1{\rm sin}}(u,v)
\right\}\,,
\eea
where we have used the shorthand notation
\bea
\ch=\cosh v\,,\qquad
\sh=\sinh v\,,\qquad
A={\rm arctan}\left(\tanh\frac{v}{2}\right)\,.
\eea

Finally, at the 2PN level we get more involved expressions, but with a similar structure
\bea
\label{calK2PNlist}
\tilde{\mathcal K}_{\rm 2PN}^{\rm LO}(u)&=&C_{K_0K_0}^{\rm 2PN, LO}f_{K_0K_0}(u)+C_{K_0K_1}^{\rm 2PN, LO}f_{K_0K_1}(u)+C_{K_1K_1}^{\rm 2PN, LO}f_{K_1K_1}(u)\nonumber\\
&&
+{\mathcal I}_{\rm 2PN}^{\rm LO}(u)
\,,\nonumber\\
\tilde{\mathcal K}_{\rm 2PN}^{\rm NLO}(u)&=&C_{K_0K_0}^{\rm 2PN, NLO}f_{K_0K_0}(u)+C_{K_0K_1}^{\rm 2PN, NLO}f_{K_0K_1}(u)+C_{K_1K_1}^{\rm 2PN, NLO}f_{K_1K_1}(u)
\nonumber\\
&&
+{\mathcal I}_{\rm 2PN}^{\rm NLO}(u)
\,,\nonumber\\
\tilde{\mathcal K}_{\rm 2PN}^{\rm NNLO}(u)&=&C_{K_0K_0}^{\rm 2PN, NNLO}f_{K_0K_0}(u)+C_{K_0K_1}^{\rm 2PN, NNLO}f_{K_0K_1}(u)+C_{K_1K_1}^{\rm 2PN, NNLO}f_{K_1K_1}(u)\nonumber\\
&&
+C_{K_0\partial^2K_0}^{\rm 2PN, NNLO}f_{K_0\partial^2K_0}(u)+C_{K_1\partial^2K_0}^{\rm 2PN, NNLO}f_{K_1\partial^2K_0}(u)
+C_{K_0\partial^2K_1}^{\rm 2PN, NNLO}f_{K_0\partial^2K_1}(u)+C_{K_1\partial^2K_1}^{\rm 2PN, NNLO}f_{K_1\partial^2K_1}(u)\nonumber\\
&&
+{\mathcal I}_{\rm 2PN}^{\rm NNLO}(u)
\,,
\eea
with
\bea
\label{calI2PNlist}
{\mathcal I}_{\rm 2PN}^{\rm LO}(u)&=&\int_{-\infty}^{\infty}dv \left[C_{K_0{\rm cos}}^{\rm 2PN, LO}(v)f_{K_0{\rm cos}}(u,v)+C_{K_1{\rm cos}}^{\rm 2PN, LO}(v)f_{K_1{\rm cos}}(u,v)\right.\nonumber\\
&&\left.
+C_{K_0{\rm sin}}^{\rm 2PN, LO}(v)f_{K_0{\rm sin}}(u,v)+C_{K_1{\rm sin}}^{\rm 2PN, LO}(v)f_{K_1{\rm sin}}(u,v)\right]
\,,\nonumber\\
{\mathcal I}_{\rm 2PN}^{\rm NLO}(u)&=&\int_{-\infty}^{\infty}dv \left[C_{K_0{\rm cos}}^{\rm 2PN, NLO}(v)f_{K_0{\rm cos}}(u,v)+C_{K_1{\rm cos}}^{\rm 2PN, NLO}(v)f_{K_1{\rm cos}}(u,v)\right.\nonumber\\
&&\left.
+C_{K_0{\rm sin}}^{\rm 2PN, NLO}(v)f_{K_0{\rm sin}}(u,v)+C_{K_1{\rm sin}}^{\rm 2PN, NLO}(v)f_{K_1{\rm sin}}(u,v)\right]
\,,\nonumber\\
{\mathcal I}_{\rm 2PN}^{\rm NNLO}(u)&=&\int_{-\infty}^{\infty}dv \left[C_{K_0{\rm cos}}^{\rm 2PN, NNLO}(v)f_{K_0{\rm cos}}(u,v)+C_{K_1{\rm cos}}^{\rm 2PN, NNLO}(v)f_{K_1{\rm cos}}(u,v)\right.\nonumber\\
&&
+C_{K_0{\rm sin}}^{\rm 2PN, NNLO}(v)f_{K_0{\rm sin}}(u,v)+C_{K_1{\rm sin}}^{\rm 2PN, NNLO}(v)f_{K_1{\rm sin}}(u,v)\nonumber\\
&&
+C_{\partial^2K_0{\rm cos}}^{\rm 2PN, NNLO}(v)f_{\partial^2K_0{\rm cos}}(u,v)+C_{\partial^2K_1{\rm cos}}^{\rm 2PN, NNLO}(v)f_{\partial^2K_1{\rm cos}}(u,v)\nonumber\\
&&\left.
+C_{\partial^2K_0{\rm sin}}^{\rm 2PN, NNLO}(v)f_{\partial^2K_0{\rm sin}}(u,v)+C_{\partial^2K_1{\rm sin}}^{\rm 2PN, NNLO}(v)f_{\partial^2K_1{\rm sin}}(u,v)
\right]\nonumber\\
&&
+\frac{72}{5}u^6\int_{-\infty}^{\infty} dv\int_{-\infty}^{\infty} dv'\, {\rm arctan}\left(\tanh\frac{v}{2}\right){\rm arctan}\left(\tanh\frac{v'}{2}\right)
e^{iu(\sinh v-\sinh v')}\nonumber\\
&&\times
\cosh v\cosh v'[\cosh^2v\cosh^2v'-2(\sinh v -\sinh v')^2]
\,,
\eea
where $C_{K_\mu K_\nu}^{\rm 2PN, LO}=\sum_ku^kC_{K_\mu K_\nu}^{\rm 2PN, LO}{}_k$, etc.
The various coefficients are listed in Tables \ref{2PNcoeffs1}--\ref{2PNcoeffs6}.

\end{widetext}


\begin{table*}
\caption{\label{2PNcoeffs1}
2PN $C_{K_\mu K_\nu}$ coefficients.
}
\begin{ruledtabular}
\begin{tabular}{c|c|c|c}
Coefficient & $C_{K_0K_0}^{\rm 2PN, LO}$ & $C_{K_0K_1}^{\rm 2PN, LO}$ & $C_{K_1K_1}^{\rm 2PN, LO}$ \\
\hline
$u^2$ & $\frac{128 \nu ^2}{315}-\frac{64 \nu }{35}+\frac{1696}{105}$ & - & $\frac{256 \nu ^2}{105}-\frac{512 \nu}{21}+\frac{8312}{315}$ \\
$u^3$ & - & $\frac{896 \nu ^2}{135}-\frac{12896 \nu}{315}+\frac{944}{135}$ & - \\
$u^4$ & $\frac{6112 \nu ^2}{945}+\frac{608 \nu}{315}-\frac{8744}{945}$ & - & $\frac{11936 \nu ^2}{945}-\frac{19696\nu }{945}+\frac{2272}{81}$ \\
$u^5$ & - & $\frac{6848\nu ^2}{945}-\frac{1376 \nu}{189}-\frac{33232}{2835}$ & - \\
$u^6$ & $\frac{328\nu ^2}{105}+\frac{4352 \nu}{945}+\frac{6824}{2835}$ & - & $\frac{872\nu ^2}{945}+\frac{1312 \nu}{945}+\frac{9544}{2835}$ \\
$u^7$ & - & $-\frac{928 \nu ^2}{189}-\frac{128 \nu}{21}+\frac{352}{189}$ & - \\
$u^8$ & $\frac{32 \nu ^2}{63}-\frac{64 \nu}{189}+\frac{32}{567}$ & - & $\frac{32 \nu ^2}{63}-\frac{64 \nu}{189}+\frac{32}{567}$ \\
\hline
Coefficient & $C_{K_0K_0}^{\rm 2PN, NLO}$ & $C_{K_0K_1}^{\rm 2PN, NLO}$ & $C_{K_1K_1}^{\rm 2PN, NLO}$ \\
\hline
$u^3$ & $\frac{128 \nu ^2}{315}-\frac{496 \nu }{105}+\frac{4432}{105}$ & - & $\frac{64 \nu ^2}{105}-\frac{184 \nu }{105}-\frac{1312}{45}$ \\
$u^4$ & - & $\frac{3104 \nu ^2}{945}-\frac{4064 \nu }{315}+\frac{3536}{135}$ & - \\
$u^5$ & $\frac{704 \nu ^2}{189}+\frac{8108 \nu }{315}-\frac{1292}{945}$ & - & $\frac{6896 \nu ^2}{945}+\frac{27752 \nu }{945}+\frac{23144}{2835}$ \\
$u^6$ & - & $\frac{3392 \nu ^2}{945}+\frac{29696 \nu }{945}-\frac{180976}{2835}$ & - \\
$u^7$ & $\frac{8 \nu ^2}{5}+\frac{17672 \nu }{945}-\frac{2896}{2835}$ & - & $-\frac{568 \nu ^2}{945}+\frac{14632 \nu }{945}-\frac{176}{2835}$ \\
$u^8$ & - & $-\frac{928 \nu ^2}{189}-\frac{128 \nu }{21}+\frac{352}{189}$ & - \\
$u^9$ & $\frac{32 \nu ^2}{63}-\frac{64 \nu }{189}+\frac{32}{567}$ & - & $\frac{32 \nu ^2}{63}-\frac{64 \nu }{189}+\frac{32}{567}$ \\
\hline
Coefficient & $C_{K_0K_0}^{\rm 2PN, NNLO}$ & $C_{K_0K_1}^{\rm 2PN, NNLO}$ & $C_{K_1K_1}^{\rm 2PN, NNLO}$ \\
\hline
$u^2$ & $\frac{128 \nu ^2}{105}+\frac{832 \nu }{105}-\frac{14144}{315}$ & - & $\frac{128 \nu ^2}{105}+\frac{496 \nu }{7}-\frac{4328}{105}$ \\
$u^3$ & - & $\frac{1184 \nu ^2}{63}+\frac{16 \nu }{45}+\frac{107056}{945}$ & - \\
$u^4$ & $\frac{8192 \nu ^2}{945}-\frac{36536 \nu }{315}+\frac{5816}{945}$ & - & $\frac{128 \nu ^2}{7}-\frac{25904 \nu }{945}-\frac{118232}{2835}$ \\
 & $+\pi ^2 \left(\frac{106 \nu ^2}{315}-\frac{652 \nu}{105}+\frac{674}{15}\right)$ & & $+\pi ^2 \left(-\frac{22 \nu ^2}{105}+\frac{68 \nu}{21}-\frac{3134}{315}\right)$ \\
$u^5$ & - & $\frac{368 \nu ^2}{15}-\frac{2576 \nu }{15}+\frac{4992}{35}$ & - \\
	& & $+\pi ^2 \left(\frac{1102 \nu ^2}{945}-\frac{884 \nu}{63}+\frac{16186}{135}\right)$ & \\
$u^6$ & $\frac{2944 \nu ^2}{315}-\frac{62128 \nu }{945}+\frac{272}{405}$ & - & $\frac{1096 \nu ^2}{105}-\frac{18496 \nu }{945}-\frac{20632}{2835}$ \\
 & $+\pi ^2 \left(\frac{842 \nu ^2}{945}+\frac{5536 \nu
   }{315}+\frac{4814}{135}\right)$ & & $+\pi ^2 \left(\frac{1306 \nu ^2}{945}+\frac{30796 \nu}{945}+\frac{75238}{2835}\right)$\\
$u^7$ & - & $\frac{2336 \nu ^2}{189}+\frac{2432 \nu }{63}-\frac{32}{3}$ & - \\
 & & $+\pi ^2 \left(-\frac{32 \nu ^2}{945}+\frac{33136 \nu}{945}-\frac{4696}{81}\right)$ & \\
$u^8$ & $-\frac{32 \nu ^2}{9}+\frac{64 \nu }{27}-\frac{32}{81}$ & - & $-\frac{64 \nu ^2}{21}+\frac{128 \nu }{63}-\frac{64}{189}$ \\
 & $+\pi ^2 \left(\frac{4 \nu ^2}{105}+\frac{15496 \nu}{945}-\frac{6308}{2835}\right)$ & & $+\pi ^2 \left(-\frac{1004 \nu ^2}{945}+\frac{13976 \nu}{945}-\frac{4948}{2835}\right)$ \\
$u^9$ & - & $\pi ^2 \left(-\frac{464 \nu ^2}{189}-\frac{64 \nu }{21}+\frac{176}{189}\right)$ & - \\
$u^{10}$ & $\pi ^2 \left(\frac{16 \nu ^2}{63}-\frac{32 \nu }{189}+\frac{16}{567}\right)$ & - & $\pi ^2 \left(\frac{16 \nu ^2}{63}-\frac{32 \nu }{189}+\frac{16}{567}\right)$ \\
\end{tabular}
\end{ruledtabular}
\end{table*}


\begin{table*}
\caption{\label{2PNcoeffs2}
2PN $C_{K_0{\rm cos}}$ coefficients.
}
\begin{ruledtabular}
\begin{tabular}{c|c}
Coefficient & $C_{K_0{\rm cos}}^{\rm 2PN, LO}$ \\
\hline
$u^4$ &
$\ch^2 \left(\frac{44 \nu ^2}{945}-\frac{820 \nu }{189}-\frac{1016}{945}\right)-\frac{526 \nu ^2}{945}+\frac{158 \nu}{189}-\frac{352}{189}+\frac{-\frac{ 613 \nu ^2}{945}+\frac{2011 \nu}{945}+\frac{262}{945}}{\ch^2}$\\
$u^5$ &
$\pi  \ch^3 \left(-\frac{191 \nu ^2}{63}+\frac{811 \nu}{315}-\frac{116}{315}\right)+\pi  \ch \left(\frac{33 \nu ^2}{14}-\frac{491 \nu}{210}+\frac{26}{105}\right)$
\\
$u^6$ &
$-\frac{128}{35} \sh v \left(\ch^3 \left(\nu ^2+\frac{4 \nu}{3}+\frac{115}{192}\right)+\frac{133}{144}\ch \left(\nu ^2+\frac{781 \nu}{266}+\frac{1579}{1064}\right)\right)-\frac{128}{35} \sh \left(3 \ch^2\left(\nu -\frac{11}{48}\right) \left(\nu+\frac{1}{4}\right) \sh+\frac{73}{64}\left(\nu ^2+\frac{215 \nu}{219}+\frac{641}{438}\right)\sh\right)$
\\
$u^7$ &
$\frac{4}{15} \pi  \ch \sh^2 \left(\ch^2 \left(\nu+\frac{1}{2}\right)^2+\frac{249 \nu^2}{14}-\frac{2 \nu}{7}-\frac{47}{56}\right)$
\\
$u^8$ &
$\frac{8}{5} \ch \left(\nu +\frac{1}{2}\right)^2 \sh^3 v$
\\
\hline
Coefficient & $C_{K_0{\rm cos}}^{\rm 2PN, NLO}$ \\
\hline
$u^4$ & $\frac{\ch^3 \left(\frac{1418 \nu ^2}{105}-\frac{1696 \nu}{105}+\frac{152}{35}\right)}{\pi }+\frac{\ch \left(-\frac{4034 \nu^2}{315}+\frac{1354 \nu }{63}-\frac{604}{45}\right)}{\pi }+\frac{-\frac{1226 \nu ^2}{945}+\frac{4022 \nu }{945}+\frac{524}{945}}{\pi \ch^3}+\frac{\frac{613 \nu ^2}{945}-\frac{1439 \nu}{189}+\frac{9026}{945}}{\pi  \ch}
+\frac{8 A \ch (96 \nu +311) \sh}{35 \pi }$ 
\\
$u^5$ & $\ch^2 \left(\frac{22 \nu ^2}{945}-\frac{410 \nu}{189}-\frac{508}{945}\right)+\frac{-\frac{613 \nu ^2}{1890}+\frac{2011 \nu}{1890}+\frac{131}{945}}{\ch^2}-\frac{263 \nu ^2}{945}+\frac{79 \nu}{189}-\frac{176}{189}$
\\
$u^6$ & $\frac{\ch v^2}{\pi } \left[\ch^2 \left(\frac{463 \nu ^2}{315}+\frac{2423 \nu}{315}+\frac{3932}{315}\right)-\frac{57 \nu^2}{70}-\frac{235 \nu }{42}-\frac{2186}{105}\right]
+\frac{\sh v}{\pi} \left[\ch^2\left(\frac{292 \nu ^2}{35}+\frac{424 \nu }{21}+\frac{862}{105}\right)+\frac{\frac{1226 \nu ^2}{945}-\frac{4022 \nu }{945}-\frac{524}{945}}{\ch^2}\right.$ 
\\
& $\left.-\frac{119 \nu ^2}{135}-\frac{14479 \nu}{945}+\frac{58973}{945}\right]
+\pi \ch  \left[\ch^2 \left(-\frac{191 \nu ^2}{252}+\frac{811 \nu}{1260}-\frac{29}{315}\right)+\frac{33 \nu ^2}{56}-\frac{491\nu }{840}+\frac{13}{210}\right]$ 
\\
& $+\frac{\ch}{\pi }\left[-\frac{8}{5} \ch^4\left(\nu +\frac{1}{2}\right)^2+\ch^2 \left(-\frac{354 \nu^2}{35}+\frac{386 \nu }{21}+\frac{86}{105}\right)+\frac{82 \nu^2}{7}-\frac{1762 \nu }{105}-\frac{44}{105}\right]
+\frac{A \sh \left(\frac{8}{105} \ch^3 (172 \nu +146)+\frac{8}{105} \ch (27-888 \nu )\right)}{\pi }$
\\
$u^7$ & $\ch\sh v \left[\ch^2 \left(-\frac{64 \nu ^2}{35}-\frac{256 \nu}{105}-\frac{23}{21}\right)-\frac{76 \nu ^2}{45}-\frac{1562\nu }{315}-\frac{1579}{630}\right]+\sh^2 \left(-\frac{192}{35} \ch^2 \left(\nu -\frac{11}{48}\right)\left(\nu +\frac{1}{4}\right)-\frac{73 \nu ^2}{35}-\frac{43 \nu}{21}-\frac{641}{210}\right)$
\\
$u^8$ & $\frac{\sh^2 v^2 \left[\ch \left(-\frac{22 \nu ^2}{7}-\frac{284 \nu}{21}-\frac{293}{42}\right)-\frac{4}{15} \ch^3 \left(\nu+\frac{1}{2}\right)^2\right]}{\pi }-\frac{2 \left(\ch^2+1\right) (2 \nu +1)^2 \sh^3 v}{5 \pi }+\pi\sh^2\ch\left[\frac{1}{15} \ch^2 \left(\nu +\frac{1}{2}\right)^2+\frac{83 \nu ^2}{70}-\frac{2 \nu}{105}-\frac{47}{840}\right]$ 
\\
$u^9$ & $\frac{4}{5} \ch \left(\nu +\frac{1}{2}\right)^2 \sh^3 v$ 
\\
\hline
Coefficient& $C_{K_0{\rm cos}}^{\rm 2PN, NNLO}$ \\
\hline
$u^4$ & $\frac{-\frac{613 \nu ^2}{315}+\frac{2011 \nu }{315}+\frac{262}{315}}{\ch^4}+\ch^2 \left(\frac{76 \nu ^2}{45}+\frac{1066 \nu
   }{63}+\frac{7052}{315}\right)+\frac{\frac{1226 \nu ^2}{315}-\frac{1150 \nu }{63}+\frac{2572}{315}}{\ch^2}+\frac{106 \nu ^2}{45}-\frac{734 \nu }{63}-\frac{1292}{63}$
\\
 & $+A \sh \left(-\frac{8}{35} \ch^2 (114 \nu +291)+\frac{432 \nu }{35}-\frac{1048}{7}\right)$
\\
$u^5$ & $\pi  \ch^3 \left(\frac{709 \nu ^2}{105}-\frac{848 \nu }{105}+\frac{76}{35}\right)+\frac{\pi  \left(-\frac{613 \nu ^2}{945}+\frac{2011 \nu}{945}+\frac{262}{945}\right)}{\ch^3}+\pi  \ch \left(-\frac{2017 \nu ^2}{315}+\frac{677 \nu }{63}-\frac{302}{45}\right)+\frac{\pi\left(\frac{613 \nu    ^2}{1890}-\frac{1439 \nu }{378}+\frac{4513}{945}\right)}{\ch}$
\\
 & $+\frac{216}{35} \pi  A \ch \left(\nu +\frac{689}{54}\right) \sh$
\\
$u^6$ & $v^2 \left(\ch^2 \left(-\frac{62 \nu ^2}{189}+\frac{6586 \nu }{945}-\frac{16988}{945}\right)+\frac{\frac{613 \nu ^2}{1890}-\frac{2011 \nu}{1890}-\frac{131}{945}}{\ch^2}+\frac{31 \nu ^2}{189}-\frac{8144 \nu }{945}+\frac{48184}{945}\right)$
\\
 & $+\sh v \left(\ch^3 \left(\frac{328 \nu ^2}{35}+\frac{212 \nu }{21}+\frac{914}{105}\right)+\frac{\frac{2452 \nu ^2}{945}-\frac{8044 \nu}{945}-\frac{1048}{945}}{\ch^3}+\ch \left(-\frac{38 \nu ^2}{21}+\frac{1144 \nu }{105}-\frac{484}{21}\right)+\frac{\frac{384 \nu}{35}-\frac{688}{35}}{\ch}\right)$
\\
 & $+\pi ^2 \left(\ch^2 \left(\frac{11 \nu ^2}{1890}-\frac{205 \nu }{378}-\frac{127}{945}\right)+\frac{-\frac{613 \nu ^2}{7560}+\frac{2011 \nu
   }{7560}+\frac{131}{3780}}{\ch^2}-\frac{263 \nu ^2}{3780}+\frac{79 \nu }{756}-\frac{44}{189}\right)$
\\
 & $+\ch^4 \left(\frac{712 \nu ^2}{35}-\frac{524 \nu }{105}+\frac{1124}{105}\right)+\ch^2 \left(-\frac{212 \nu ^2}{15}-\frac{3484 \nu }{105}+\frac{44}{5}\right)-\frac{652\nu ^2}{105}+\frac{1336 \nu }{35}-\frac{2048}{105}$
\\
 & $+A \left(v \left(\frac{2896}{105} \ch \left(\nu +\frac{3289}{362}\right)-\frac{368}{35} \ch^3 \left(\nu +\frac{531}{46}\right)\right)+\sh \left(\frac{1536}{35}\ch^2 \left(\nu -\frac{23}{96}\right)+\frac{2368 \nu }{35}-\frac{72}{35}\right)\right)$
\\
$u^7$ & $\pi  v^2 \left(\ch^3 \left(\frac{463 \nu ^2}{630}+\frac{2423 \nu }{630}+\frac{1966}{315}\right)+\ch \left(-\frac{57 \nu ^2}{140}-\frac{235 \nu}{84}-\frac{1093}{105}\right)\right)$
\\
 & $+\sh v \left(\pi  \ch^2 \left(\frac{146 \nu ^2}{35}+\frac{212 \nu }{21}+\frac{431}{105}\right)+\frac{\pi  \left(\frac{613 \nu^2}{945}-\frac{2011 \nu }{945}-\frac{262}{945}\right)}{\ch^2}+\pi  \left(-\frac{119 \nu ^2}{270}-\frac{14479 \nu }{1890}+\frac{58973}{1890}\right)\right)$
\\
 & $+\pi ^3 \left(\ch^3 \left(-\frac{191 \nu ^2}{1512}+\frac{811 \nu }{7560}-\frac{29}{1890}\right)+\ch \left(\frac{11 \nu ^2}{112}-\frac{491 \nu}{5040}+\frac{13}{1260}\right)\right)
+\pi  A \sh \left(\frac{688}{105} \ch^3 \left(\nu +\frac{73}{86}\right)-\frac{1184}{35} \ch \left(\nu -\frac{9}{296}\right)\right)$
\\
 & $+\pi  \left(-\frac{4}{5} \ch^5 \left(\nu +\frac{1}{2}\right)^2+\ch^3 \left(-\frac{177 \nu ^2}{35}+\frac{193 \nu
   }{21}+\frac{43}{105}\right)+\ch \left(\frac{41 \nu ^2}{7}-\frac{881 \nu }{105}-\frac{22}{105}\right)\right)$
\\
$u^8$ & $\sh v^3 \left(\ch^3 \left(\frac{136 \nu ^2}{315}+\frac{244 \nu }{105}+\frac{367}{315}\right)+\ch \left(-\frac{16 \nu ^2}{189}+\frac{1226 \nu}{945}+\frac{9313}{270}\right)\right)+\sh^2 v^2 \left(\ch^2 \left(\frac{136 \nu ^2}{35}+\frac{96 \nu }{7}+\frac{241}{35}\right)+\frac{17 \nu ^2}{35}+\frac{1643 \nu}{105}+\frac{2153}{210}\right)$
\\
 & $+\sh v \left(\ch^3 \left(-\frac{8 \nu ^2}{5}+\pi ^2 \left(-\frac{16 \nu ^2}{35}-\frac{64 \nu }{105}-\frac{23}{84}\right)-\frac{8 \nu }{5}-\frac{2}{5}\right)+\ch\left(\frac{8 \nu ^2}{5}+\pi ^2 \left(-\frac{19 \nu ^2}{45}-\frac{781 \nu }{630}-\frac{1579}{2520}\right)+\frac{8 \nu }{5}+\frac{2}{5}\right)\right)$
\\
 & $+\pi ^2 \sh^2 \left(-\frac{48}{35} \ch^2 \left(\nu -\frac{11}{48}\right) \left(\nu +\frac{1}{4}\right)-\frac{73 \nu ^2}{140}-\frac{43 \nu }{84}-\frac{641}{840}\right)
-\frac{8}{7} A \ch \left(3 \ch^2-4\right) (4 \nu -1) v$
\\
$u^9$ & $\pi  \sh^2 v^2 \left(-\frac{2}{15} \ch^3 \left(\nu +\frac{1}{2}\right)^2-\frac{2}{15} \ch \left(\frac{165 \nu ^2}{14}+\frac{355 \nu}{7}+\frac{1465}{56}\right)\right)-\frac{1}{5} \pi  \left(\ch^2+1\right) (2 \nu +1)^2 \sh^3 v$
\\
& $+\pi ^3 \sh^2 \left(\frac{1}{90} \ch^3 \left(\nu +\frac{1}{2}\right)^2-\frac{2}{15} \ch \left(-\frac{83 \nu^2}{56}+\frac{\nu }{42}+\frac{47}{672}\right)\right)$
\\
$u^{10}$ & $\frac{1}{60} \ch (2 \nu +1)^2 \sh^3 v \left(3 \pi ^2-4 v^2\right)$
\\
\end{tabular}
\end{ruledtabular}
\end{table*}


\begin{table*}
\caption{\label{2PNcoeffs3}
2PN $C_{K_1{\rm cos}}$ coefficients.
}
\begin{ruledtabular}
\begin{tabular}{c|c}
Coefficient & $C_{K_1{\rm cos}}^{\rm 2PN, LO}$ \\
\hline
$u^5$ & $\ch^4 \left(-\frac{384 \nu^2}{35}-\frac{8 \nu}{35}+\frac{22}{35}\right)+\ch^2\left(\frac{256 \nu ^2}{105}-\frac{256 \nu}{35}-\frac{64}{7}\right)+\frac{-\frac{1226 \nu ^2}{945}+\frac{4022 \nu }{945}+\frac{524}{945}}{\ch^2}+\frac{373 \nu ^2}{105}+\frac{97 \nu}{21}+\frac{316}{105}$
\\
& $-\frac{2288}{315} \ch \left(\nu^2+\frac{334 \nu}{143}+\frac{1349}{1144}\right) \sh v$
\\
$u^6$ & $\pi\ch^3 \left(\frac{25 \nu ^2}{42}+\frac{293 \nu}{70}-\frac{71}{105}\right)+\pi\ch\left(\frac{7 \nu ^2}{9}-\frac{367 \nu}{63}+\frac{247}{315}\right)$ 
\\
$u^7$ & $-\frac{136}{35} \sh v\left(\ch^3 \left(\nu ^2+\frac{25 \nu}{17}+\frac{101}{136}\right)-\frac{31}{17}\ch \left(\nu ^2+\frac{166 \nu}{93}+\frac{86}{93}\right)\right)+\frac{4}{5} \left(\ch^2+2\right)\left(\nu +\frac{1}{2}\right)^2\sh^2$
\\
$u^8$ & $\frac{2}{5} \pi  \ch \left(\ch^2-2\right) \left(\nu+\frac{1}{2}\right)^2 \sh^2$  
\\
\hline
Coefficient & $C_{K_1{\rm cos}}^{\rm 2PN, NLO}$ \\
\hline
$u^5$ & $\frac{\sh v}{\pi } \left[\ch^2 \left(\frac{926 \nu ^2}{105}+\frac{386 \nu}{21}+\frac{646}{105}\right)+\frac{328 \nu^2}{945}-\frac{5812 \nu }{945}+\frac{65426}{945}+\frac{\frac{1226 \nu ^2}{945}-\frac{4022\nu }{945}-\frac{524}{945}}{\ch^2}\right]+\frac{\ch^3}{\pi } \left(\frac{521 \nu ^2}{35}-\frac{241 \nu}{35}+\frac{44}{7}\right)$
\\
& $+\frac{\ch}{\pi } \left(-\frac{11554 \nu ^2}{945}+\frac{238 \nu}{135}-\frac{17858}{945}\right)+\frac{\frac{1226 \nu^2}{945}-\frac{2878 \nu }{189}+\frac{18052}{945}}{\pi \ch}+\frac{-\frac{2452 \nu ^2}{945}+\frac{8044 \nu }{945}+\frac{1048}{945}}{\pi \ch^3}
+\frac{A \sh \ch\left[\frac{16}{105} \ch^2 (78 \nu +75)+\frac{16}{105}  (130 \nu +506)\right]}{\pi }$ 
\\
$u^6$ & $\sh\ch v \left(-\frac{1144 \nu^2}{315}-\frac{2672 \nu }{315}-\frac{1349}{315}\right) + \ch^4 \left(-\frac{192 \nu ^2}{35}-\frac{4 \nu}{35}+\frac{11}{35}\right)+\ch^2 \left(\frac{128 \nu ^2}{105}-\frac{128\nu }{35}-\frac{32}{7}\right)$
\\
& $+\frac{-\frac{613 \nu ^2}{945}+\frac{2011 \nu}{945}+\frac{262}{945}}{\ch^2}+\frac{373\nu ^2}{210}+\frac{97 \nu }{42}+\frac{158}{105}$
\\
$u^7$ & $\frac{\ch v^2}{\pi } \left[\ch^2 \left(-\frac{281 \nu ^2}{210}-\frac{167 \nu}{70}+\frac{1223}{105}\right)+\frac{367 \nu ^2}{315}-\frac{223\nu }{315}-\frac{1501}{45}\right]+\frac{\sh v}{\pi } \left[-\frac{8}{5} \ch^4 \left(\nu+\frac{1}{2}\right)^2+\ch^2 \left(-\frac{149 \nu ^2}{35}-\frac{421 \nu}{105}-\frac{41}{105}\right)-\frac{62 \nu ^2}{35}\right.$
\\
& $\left.-\frac{1774 \nu}{105}-\frac{1376}{105}\right]
+\pi \ch  \left[\ch^2 \left(\frac{25 \nu ^2}{168}+\frac{293 \nu}{280}-\frac{71}{420}\right)+\frac{7 \nu ^2}{36}-\frac{367 \nu}{252}+\frac{247}{1260}\right]-\frac{\ch \sh^2 \left(3 \ch^2-1\right) (2 \nu +1)^2}{5 \pi }
-\frac{8 A \ch \left(\ch^2-4\right) (4 \nu -1) \sh}{7 \pi }$
\\
$u^8$ & $\ch \sh v \left[\ch^2 \left(-\frac{68 \nu ^2}{35}-\frac{20 \nu}{7}-\frac{101}{70}\right)+\frac{124 \nu ^2}{35}+\frac{664 \nu}{105}+\frac{344}{105}\right]+\frac{1}{10} \left(\ch^2+2\right) (2 \nu +1)^2 \sh^2$
\\
$u^9$ & $\frac{\ch \left(\ch^2-2\right) (2 \nu +1)^2 \sh^2 \left(\pi^2-4 v^2\right)}{40 \pi }$ 
\\
\hline
Coefficient& $C_{K_1{\rm cos}}^{\rm 2PN, NNLO}$ \\
\hline
$u^5$ & $\sh v\left(\frac{\frac{2452 \nu ^2}{945}-\frac{8044 \nu }{945}-\frac{1048}{945}}{\ch^3}+\ch \left(\frac{244 \nu ^2}{315}+\frac{278 \nu}{315}-\frac{7444}{315}\right)+\frac{\frac{384 \nu }{35}-\frac{688}{35}}{\ch}\right)+\ch^4 \left(\frac{328 \nu ^2}{35}-\frac{204 \nu }{35}+\frac{402}{35}\right)$
\\
& $+\frac{-\frac{1226 \nu ^2}{315}+\frac{4022 \nu }{315}+\frac{524}{315}}{\ch^4}+\ch^2 \left(-\frac{1948 \nu ^2}{315}+\frac{2692 \nu
   }{315}+\frac{5224}{315}\right)+\frac{\frac{1226 \nu ^2}{189}-\frac{4354 \nu }{135}+\frac{15956}{945}}{\ch^2}-\frac{5 \nu ^2}{63}+\frac{3181 \nu }{315}-\frac{10064}{315}$
	\\
& $+A \left(v \left(\frac{864}{35} \ch \left(\nu +\frac{311}{54}\right)-\frac{912}{35} \ch^3 \left(\nu +\frac{97}{38}\right)\right)+\sh \left(-\frac{1248}{35}\ch^2 \left(\nu +\frac{88}{39}\right)+\frac{1024 \nu }{21}-\frac{24608}{105}\right)\right)$
\\
$u^6$ & $\sh v \left(\pi  \ch^2 \left(\frac{463 \nu ^2}{105}+\frac{193 \nu }{21}+\frac{323}{105}\right)+\frac{\pi  \left(\frac{613 \nu ^2}{945}-\frac{2011 \nu}{945}-\frac{262}{945}\right)}{\ch^2}+\pi  \left(\frac{164 \nu ^2}{945}-\frac{2906 \nu }{945}+\frac{32713}{945}\right)\right)$
\\
& $+\pi  \ch^3 \left(\frac{521 \nu ^2}{70}-\frac{241 \nu }{70}+\frac{22}{7}\right)+\frac{\pi  \left(-\frac{1226 \nu ^2}{945}+\frac{4022 \nu
   }{945}+\frac{524}{945}\right)}{\ch^3}+\pi  \ch \left(-\frac{5777 \nu ^2}{945}+\frac{119 \nu }{135}-\frac{8929}{945}\right)+\frac{\pi  \left(\frac{613 \nu
   ^2}{945}-\frac{1439 \nu }{189}+\frac{9026}{945}\right)}{\ch}$
\\
& $+\pi  A \sh \left(\frac{208}{35} \ch^3 \left(\nu +\frac{25}{26}\right)+\frac{32}{105} \ch (\nu +410)\right)$
\\
$u^7$ & $v^2 \left(\ch^4 \left(\frac{136 \nu ^2}{35}+\frac{96 \nu }{7}+\frac{241}{35}\right)+\ch^2 \left(-\frac{64 \nu ^2}{21}+\frac{1819 \nu}{105}-\frac{116}{35}\right)+\frac{\frac{613 \nu ^2}{945}-\frac{2011 \nu }{945}-\frac{262}{945}}{\ch^2}-\frac{181 \nu ^2}{210}-\frac{1963 \nu}{70}+\frac{9454}{105}\right)$
\\
 & $+\sh v \left(\ch^3 \left(\frac{464 \nu ^2}{35}+\frac{516 \nu }{35}+\frac{457}{35}\right)+\ch \left(-\frac{54 \nu ^2}{5}+\pi ^2 \left(-\frac{286 \nu^2}{315}-\frac{668 \nu }{315}-\frac{1349}{1260}\right)+\frac{122 \nu }{15}-\frac{47}{15}\right)\right)$
\\
 & $+\pi ^2 \left(\ch^4 \left(-\frac{48 \nu ^2}{35}-\frac{\nu }{35}+\frac{11}{140}\right)+\ch^2 \left(\frac{32 \nu ^2}{105}-\frac{32 \nu
   }{35}-\frac{8}{7}\right)+\frac{-\frac{613 \nu ^2}{3780}+\frac{2011 \nu }{3780}+\frac{131}{1890}}{\ch^2}+\frac{373 \nu ^2}{840}+\frac{97 \nu }{168}+\frac{79}{210}\right)$
\\
& $+\ch^4 \left(\frac{4 \nu ^2}{5}-\frac{52 \nu }{35}+\frac{27}{35}\right)+\ch^2 \left(-\frac{16 \nu ^2}{5}+\frac{288 \nu }{35}-\frac{128}{35}\right)+\frac{12 \nu^2}{5}-\frac{236 \nu }{35}+\frac{101}{35}+\ch \left(\frac{8 \nu ^2}{27}+\frac{4478 \nu }{945}+\frac{34289}{945}\right) \sh v^3$
\\
& $+A \left(v \left(\frac{4208}{105} \ch^3 \left(\nu +\frac{155}{526}\right)-\frac{2048}{35} \ch \left(\nu +\frac{1}{256}\right)\right)-\frac{8}{7} \left(5
   \ch^2+4\right) (4 \nu -1) \sh\right)$
\\
$u^8$ & $\pi  v^2 \left(-\frac{281}{420} \ch^3 \left(\nu ^2+\frac{501 \nu }{281}-\frac{2446}{281}\right)-\frac{281}{420} \ch \left(-\frac{734 \nu ^2}{843}+\frac{446 \nu }{843}+\frac{21014}{843}\right)\right)$
\\
 & $+\sh v \left(-\frac{4}{5} \pi  \ch^4 \left(\nu +\frac{1}{2}\right)^2-\frac{149}{70} \pi  \ch^2 \left(\nu ^2+\frac{421\nu }{447}+\frac{41}{447}\right)-\frac{31}{35} \pi  \left(\nu ^2+\frac{887 \nu }{93}+\frac{688}{93}\right)\right)$
\\
 & $+\pi ^3 \left(-\frac{281}{420} \ch^3 \left(-\frac{125 \nu ^2}{3372}-\frac{293 \nu }{1124}+\frac{71}{1686}\right)-\frac{281}{420} \ch \left(-\frac{245 \nu^2}{5058}+\frac{1835 \nu }{5058}-\frac{247}{5058}\right)\right)
-\frac{4}{7} \pi  A \ch \left(\ch^2-4\right) (4 \nu -1) \sh$
\\
 & $+\pi  \left(-\frac{6}{5} \ch^5 \left(\nu +\frac{1}{2}\right)^2-\frac{281}{420} \ch^3\left(-\frac{672 \nu ^2}{281}-\frac{672 \nu }{281}-\frac{168}{281}\right)-\frac{281}{420} \ch \left(\frac{168 \nu ^2}{281}+\frac{168 \nu}{281}+\frac{42}{281}\right)\right)$
\\
$u^9$ & $\sh v^3 \left(\ch^3 \left(\frac{8 \nu ^2}{21}+\frac{338 \nu }{105}+\frac{353}{210}\right)+\ch \left(-\frac{68 \nu ^2}{105}-\frac{2092 \nu}{315}-\frac{220}{63}\right)\right)+\frac{1}{40} \left(\ch^2+2\right) (2 \nu +1)^2 \sh^2 \left(\pi ^2-4 v^2\right)$
\\
& $+\sh v \left(\frac{31}{35} \pi ^2 \ch \left(\nu ^2+\frac{166 \nu }{93}+\frac{86}{93}\right)-\frac{17}{35} \pi ^2 \ch^3
   \left(\nu ^2+\frac{25 \nu }{17}+\frac{101}{136}\right)\right)$
\\
$u^{10}$ & $\frac{1}{240} \pi  \ch \left(\ch^2-2\right) (2 \nu +1)^2 \sh^2 \left(\pi ^2-12 v^2\right)$
\\
\end{tabular}
\end{ruledtabular}
\end{table*}


\begin{table*}
\caption{\label{2PNcoeffs4}
2PN $C_{K_0{\rm sin}}$ coefficients.
}
\begin{ruledtabular}
\begin{tabular}{c|c}
Coefficient & $C_{K_0{\rm sin}}^{\rm 2PN, LO}$ \\
\hline
$u^5$ & $v \left(\ch^3 \left(-\frac{1376 \nu^2}{315}-\frac{2368 \nu}{315}-\frac{646}{315}\right)+\ch\left(\frac{104 \nu ^2}{35}+\frac{48 \nu}{7}+\frac{73}{15}\right)\right)+\ch^2 \left(-\frac{442 \nu^2}{35}+\frac{1114 \nu}{105}-\frac{64}{15}\right)\sh+\left(\frac{563 \nu^2}{945}+\frac{3391 \nu}{945}+\frac{319}{945}\right) \sh$
\\
& $+\frac{\left(-\frac{1226 \nu ^2}{945}+\frac{4022\nu }{945}+\frac{524}{945}\right)\sh}{\ch^2}$ \\
$u^6$ & $\pi  \ch^3 \left(\frac{44 \nu^2}{21}+\frac{6 \nu}{35}-\frac{11}{105}\right) \sh+\pi \ch \left(\frac{1027 \nu^2}{315}-\frac{1672 \nu}{315}+\frac{391}{630}\right) \sh$
\\
$u^7$ &
$\frac{8}{15} \sh^2 v \left(\ch^3\left(\nu+\frac{1}{2}\right)^2+\frac{207}{14}\ch \left(\nu ^2+\frac{353 \nu}{207}+\frac{709}{828}\right)\right)+\frac{8}{5} \left(\ch^2+1\right) \left(\nu+\frac{1}{2}\right)^2 \sh^3$ 
\\
$u^8$ & $-\frac{4}{5} \pi  \ch \left(\nu+\frac{1}{2}\right)^2 \sh^3$ 
\\
\hline
Coefficient & $C_{K_0{\rm sin}}^{\rm 2PN, NLO}$ \\
\hline
$u^5$ &
$v \left[\frac{\ch^2 \left(\frac{332 \nu ^2}{945}-\frac{8636 \nu }{945}+\frac{3296}{189}\right)}{\pi }+\frac{-\frac{613 \nu ^2}{945}+\frac{2011 \nu}{945}+\frac{262}{945}}{\pi  \ch^2}+\frac{-\frac{229 \nu ^2}{945}+\frac{5137 \nu }{945}-\frac{6751}{189}}{\pi }\right]
+A \ch\left[\frac{8 \ch^2 (264 \nu +459)}{105 \pi }+\frac{8  (-614 \nu -1021)}{105 \pi }\right]$  
\\
& $+\frac{\sh}{\pi}\left[\ch^3 \left(-\frac{384 \nu ^2}{35}+\frac{368 \nu }{105}-\frac{158}{105}\right)+\ch \left(-\frac{236 \nu ^2}{105}-\frac{962 \nu}{105}-\frac{1378}{105}\right) +\frac{-\frac{2452 \nu ^2}{945}+\frac{8044 \nu }{945}+\frac{1048}{945}}{\ch^3}+\frac{\frac{688}{35}-\frac{384 \nu }{35}}{\ch}\right]$  
\\
$u^6$ &$\ch v \left[\ch^2 \left(-\frac{688 \nu ^2}{315}-\frac{1184 \nu }{315}-\frac{323}{315}\right)+\frac{52 \nu ^2}{35}+\frac{24 \nu }{7}+\frac{73}{30}\right]$
\\
& $+\sh \left[\ch^2 \left(-\frac{221 \nu ^2}{35}+\frac{557 \nu }{105}-\frac{32}{15}\right)+\frac{-\frac{613 \nu ^2}{945}+\frac{2011 \nu}{945}+\frac{262}{945}}{\ch^2}+\frac{563 \nu ^2}{1890}+\frac{3391 \nu }{1890}+\frac{319}{1890}\right]$ 
\\
$u^7$ &
$-\frac{\ch \sh v^2}{\pi } \left[\ch^2 \left(\frac{164 \nu ^2}{105}+\frac{494 \nu}{105}+\frac{241}{105}\right)+\frac{163 \nu^2}{315}+\frac{2528 \nu }{315}+\frac{23179}{630}\right]-\frac{\sh^2 v}{\pi } \left[\ch^2 \left(\frac{328 \nu ^2}{35}+\frac{484 \nu}{35}+\frac{46}{7}\right)+\frac{18 \nu ^2}{7}+\frac{1858 \nu}{105}+\frac{1397}{105}\right)$ 
\\
& $+\pi  \sh\ch \left[\ch^2 \left(\frac{11 \nu ^2}{21}+\frac{3 \nu}{70}-\frac{11}{420}\right)+\frac{1027 \nu ^2}{1260}-\frac{418\nu }{315}+\frac{391}{2520}\right]+\frac{2 \ch \sh^2 (2 \nu +1)^2 \sh}{5 \pi }
+\frac{8 A \ch \left(3 \ch^2-4\right) (4 \nu -1)}{7 \pi }$ 
\\
$u^8$ &
$\ch \sh^2 v \left[\frac{4}{15} \ch^2 \left(\nu+\frac{1}{2}\right)^2+\frac{138 \nu ^2}{35}+\frac{706 \nu}{105}+\frac{709}{210}\right]+\frac{1}{5} \left(\ch^2+1\right) (2\nu +1)^2 \sh^3$
\\
$u^9$ &$-\frac{\ch (2 \nu +1)^2 \sh^3 \left(\pi ^2-4 v^2\right)}{20 \pi }$ 
\\
\hline
Coefficient& $C_{K_0{\rm sin}}^{\rm 2PN, NNLO}$ \\
\hline
$u^5$ & $v \left[\ch^3 \left(\frac{968 \nu ^2}{105}+\frac{362 \nu }{105}+\frac{514}{35}\right)+\frac{-\frac{1226 \nu ^2}{945}+\frac{4022 \nu}{945}+\frac{524}{945}}{\ch^3}+\ch \left(-\frac{2588 \nu ^2}{315}+\frac{932 \nu }{315}-\frac{3418}{315}\right)+\frac{\frac{613 \nu ^2}{945}-\frac{1439 \nu}{189}+\frac{9026}{945}}{\ch}\right]$
\\
 & $+\sh \left[\frac{-\frac{1226 \nu ^2}{315}+\frac{4022 \nu }{315}+\frac{524}{315}}{\ch^4}+\ch^2 \left(\frac{2026 \nu ^2}{105}-\frac{4612 \nu}{105}+\frac{2024}{105}\right)+\frac{\frac{4904 \nu ^2}{945}-\frac{26456 \nu }{945}+\frac{3296}{189}}{\ch^2}-\frac{1898 \nu ^2}{945}-\frac{12604 \nu}{945}-\frac{2488}{135}\right]$
\\
& $+A \left[\frac{416}{35} \ch^4 \left(\nu +\frac{25}{26}\right)+\frac{5104}{105} \ch^2 \left(\nu -\frac{587}{638}\right)+\frac{432}{35} \ch \left(\nu
   +\frac{689}{54}\right) \sh v-\frac{2288 \nu }{105}+\frac{4936}{21}\right]$
\\
$u^6$ & $\pi v\left[ \ch^2 \left(\frac{166 \nu ^2}{945}-\frac{4318 \nu }{945}+\frac{1648}{189}\right)+\frac{-\frac{613 \nu ^2}{1890}+\frac{2011 \nu
   }{1890}+\frac{131}{945}}{\ch^2}-\frac{229 \nu ^2}{1890}+\frac{5137 \nu }{1890}-\frac{6751}{378}\right]$
\\
& $+\pi \sh \left[\ch^3 \left(-\frac{192 \nu ^2}{35}+\frac{184 \nu }{105}-\frac{79}{105}\right)+\frac{-\frac{1226 \nu ^2}{945}+\frac{4022 \nu
   }{945}+\frac{524}{945}}{\ch^3}+\ch \left(-\frac{118 \nu ^2}{105}-\frac{481 \nu }{105}-\frac{689}{105}\right)+\frac{\frac{344}{35}-\frac{192\nu }{35}}{\ch}\right]$
\\
& $+\pi  A \left[\frac{184}{35} \ch^3 \left(\nu +\frac{531}{46}\right)-\frac{1448}{105} \ch \left(\nu +\frac{3289}{362}\right)\right]$
\\
$u^7$ &  \ch $v^3\left[\ch^2 \left(\frac{8 \nu ^2}{27}+\frac{2906 \nu }{945}+\frac{10943}{945}\right)-\frac{4 \nu ^2}{35}-\frac{437 \nu}{315}-\frac{2327}{126}\right]
+\sh v^2 \left[\ch^2 \left(\frac{17 \nu ^2}{7}+\frac{1921 \nu }{105}+\frac{808}{21}\right)-\frac{347 \nu ^2}{1890}-\frac{7835 \nu }{378}+\frac{25643}{270}\right.$
\\
 & $\left.+\frac{\frac{613 \nu^2}{945}-\frac{2011 \nu }{945}-\frac{262}{945}}{\ch^2}\right]
+v \left[-\frac{8}{5} \ch^5 \left(\nu +\frac{1}{2}\right)^2+\ch^3 \left(-\frac{298 \nu ^2}{35}+\pi ^2 \left(-\frac{172 \nu ^2}{315}-\frac{296 \nu}{315}-\frac{323}{1260}\right)+\frac{502 \nu }{105}-\frac{134}{21}\right)\right.$
\\
 & $\left.+\ch \left(\frac{354 \nu ^2}{35}+\pi ^2 \left(\frac{13 \nu ^2}{35}+\frac{6 \nu}{7}+\frac{73}{120}\right)-\frac{334 \nu }{105}+\frac{712}{105}\right)\right]
+\sh \left[-\frac{16}{5} \ch^4 \left(\nu+\frac{1}{2}\right)^2+\ch^2 \left(\frac{48 \nu }{7}-\frac{12}{7}\right)+\frac{16 \nu ^2}{5}-\frac{208 \nu }{35}+\frac{108}{35}\right]$
\\
 & $+\pi ^2 \sh \left(\ch^2 \left(-\frac{221 \nu ^2}{140}+\frac{557 \nu }{420}-\frac{8}{15}\right)+\frac{-\frac{613 \nu ^2}{3780}+\frac{2011 \nu   }{3780}+\frac{131}{1890}}{\ch^2}+\frac{563 \nu ^2}{7560}+\frac{3391 \nu }{7560}+\frac{319}{7560}\right)$
\\
 & $+A \left[\sh v \left(\frac{1376}{105} \ch^3 \left(\nu +\frac{73}{86}\right)-\frac{2368}{35} \ch \left(\nu-\frac{9}{296}\right)\right)-\frac{24}{7} \sh^2\ch^2 (4 \nu -1)+\frac{128 \nu }{7}-\frac{32}{7}\right]$
\\
$u^8$ & $-\pi  \sh\ch  v^2 \left[\ch^2 \left(\frac{82 \nu ^2}{105}+\frac{247 \nu }{105}+\frac{241}{210}\right)+\frac{163 \nu ^2}{630}+\frac{1264 \nu }{315}+\frac{23179}{1260}\right]-\pi  \sh^2 v \left[\ch^2 \left(\frac{164 \nu ^2}{35}+\frac{242 \nu }{35}+\frac{23}{7}\right)+\frac{9 \nu ^2}{7}+\frac{929\nu }{105}+\frac{1397}{210}\right]$
\\
& $+\pi ^3 \sh\ch \left[\ch^2 \left(\frac{11 \nu ^2}{126}+\frac{\nu }{140}-\frac{11}{2520}\right)+\frac{1027 \nu ^2}{7560}-\frac{209 \nu}{945}+\frac{391}{15120}\right]+\frac{1}{5} \pi  \ch (2 \nu +1)^2 \sh^3
+\frac{4}{7} \pi  A \ch \left(3 \ch^2-4\right) (4 \nu -1)$
\\
$u^9$ & $-v^3 \ch\sh^2 \left[\frac{82 \nu ^2}{105}+\frac{2134 \nu }{315}+\frac{2221}{630} +\frac{4}{45} \ch^2 \left(\nu +\frac{1}{2}\right)^2\right]+\frac{1}{20} \left(\ch^2+1\right) (2 \nu +1)^2 \sh^3 \left(\pi ^2-4
   v^2\right)$
\\ 
& $+\pi ^2  \ch\sh^2 v \left[\frac{1}{15} \ch^2 \left(\nu +\frac{1}{2}\right)^2+\frac{69}{70} \left(\nu ^2+\frac{353 \nu}{207}+\frac{709}{828}\right)\right]$
\\
$u^{10}$ & $-\frac{1}{120} \pi  \ch (2 \nu +1)^2 \sh^3 \left(\pi ^2-12 v^2\right)$
\\
\end{tabular}
\end{ruledtabular}
\end{table*}


\begin{table*}
\caption{\label{2PNcoeffs5}
2PN $C_{K_1{\rm sin}}$ coefficients.
}
\begin{ruledtabular}
\begin{tabular}{c|c}
Coefficient & $C_{K_1{\rm sin}}^{\rm 2PN, LO}$ \\
\hline
$u^4$ & $\ch^2 \left(-\frac{1376 \nu^2}{105}+\frac{1304 \nu}{105}-\frac{232}{105}\right)\sh+\left(-\frac{1354 \nu^2}{945}+\frac{230 \nu}{189}-\frac{2732}{945}\right) \sh+\frac{\left(-\frac{1226 \nu ^2}{945}+\frac{4022\nu }{945}+\frac{524}{945}\right)\sh}{\ch^2}$ 
\\
$u^5$ & $\pi  \ch \left(\frac{353 \nu^2}{63}-\frac{1633 \nu}{315}+\frac{188}{315}\right) \sh$ 
\\
$u^6$ & $v \left(\ch^3 \left(\frac{32 \nu^2}{15}+\frac{104 \nu}{35}+\frac{407}{105}\right)+\ch\left(-\frac{248 \nu ^2}{315}+\frac{656 \nu }{315}+\frac{548}{315}\right)\right)+\frac{8}{5} \ch^4 \left(\nu +\frac{1}{2}\right)^2 \sh+\ch^2 \left(\frac{177 \nu ^2}{35}-\frac{293 \nu}{105}-\frac{337}{105}\right)\sh$
\\
& $+\left(\frac{118 \nu^2}{35}+\frac{346 \nu}{105}+\frac{124}{21}\right) \sh$\\
$u^7$ & $\pi  \ch^3 \left(\frac{82 \nu^2}{35}-\frac{19 \nu}{35}-\frac{5}{14}\right) \sh+\pi \ch \left(-\frac{152 \nu^2}{35}+\frac{10 \nu}{21}+\frac{34}{105}\right)\sh$ 
\\
$u^8$ & $\frac{4}{5} \ch\left(\ch^2-2\right) \left(\nu+\frac{1}{2}\right)^2 \sh^2 v$  
\\
\hline
Coefficient & $C_{K_1{\rm sin}}^{\rm 2PN, NLO}$ \\
\hline
$u^4$ & $\frac{\sh}{\pi} \left[\frac{-\frac{2452 \nu ^2}{945}+\frac{8044 \nu }{945}+\frac{1048}{945}}{\ch^3}+\ch \left(-\frac{112 \nu ^2}{45}-\frac{4058 \nu
   }{315}-\frac{572}{45}\right)+\frac{\frac{688}{35}-\frac{384 \nu }{35}}{\ch}\right]
	+A \ch\left[\frac{8 \ch^2 (156 \nu -87)}{35 \pi }+\frac{8  (134-192 \nu )}{35 \pi }\right]$ \\
$u^5$ & $\sh \left[\ch^2 \left(-\frac{688 \nu ^2}{105}+\frac{652 \nu }{105}-\frac{116}{105}\right)+\frac{-\frac{613 \nu ^2}{945}+\frac{2011 \nu}{945}+\frac{262}{945}}{\ch^2}-\frac{677 \nu ^2}{945}+\frac{115 \nu }{189}-\frac{1366}{945}\right]$ 
\\
$u^6$ & $v\left[\frac{\ch^4 \left(-\frac{328 \nu ^2}{35}-\frac{484 \nu }{35}-\frac{46}{7}\right)}{\pi }+\frac{\ch^2 \left(\frac{67 \nu ^2}{15}-\frac{2581 \nu }{105}+\frac{523}{35}\right)}{\pi }+\frac{\frac{319 \nu ^2}{105}+\frac{2431 \nu}{105}-\frac{842}{15}}{\pi }+\frac{-\frac{1226 \nu ^2}{945}+\frac{4022 \nu }{945}+\frac{524}{945}}{\pi  \ch^2}\right]$
\\
& $-\frac{\ch \sh v^2}{\pi } \left(\frac{649 \nu ^2}{315}+\frac{4709 \nu }{315}+\frac{12716}{315}\right)+\frac{\sh \ch}{\pi }\left[\ch^2 \left(-\frac{548 \nu ^2}{35}+\frac{198 \nu }{35}-\frac{79}{35}\right)+\frac{58 \nu ^2}{5}-\frac{224 \nu}{15}-\frac{7}{15}\right]$
\\
& $+\pi \sh\ch \left(\frac{353 \nu ^2}{252}-\frac{1633 \nu }{1260}+\frac{47}{315}\right)
-A \ch\left[\frac{8 \ch^2 (526 \nu +155)}{105 \pi }+\frac{8  (-768 \nu -3)}{105 \pi }\right]$
\\
$u^7$ & $\ch v \left[\ch^2 \left(\frac{16 \nu ^2}{15}+\frac{52 \nu}{35}+\frac{407}{210}\right)-\frac{124 \nu ^2}{315}+\frac{328\nu }{315}+\frac{274}{315}\right]$  
\\
& $+\sh \left[\frac{4}{5} \ch^4 \left(\nu +\frac{1}{2}\right)^2+\ch^2 \left(\frac{177 \nu ^2}{70}-\frac{293 \nu }{210}-\frac{337}{210}\right)+\frac{59 \nu^2}{35}+\frac{173 \nu }{105}+\frac{62}{21}\right]$ 
\\
$u^8$ & $-\frac{\sh\ch  v^2}{\pi} \left[\frac{54 \ch^2 \left(\nu ^2+\frac{73 \nu}{18}+\frac{227}{108}\right)}{35}+\frac{54 \left(-\frac{16 \nu^2}{9}-\frac{689 \nu }{81}-\frac{361}{81}\right)}{35}\right]+\frac{\sh^2 v}{5 \pi }\left(\ch^2+2\right) (2 \nu +1)^2 $ 
\\
& $-\pi  \sh\ch \left[\frac{54}{35} \ch^2 \left(-\frac{41 \nu^2}{108}+\frac{19 \nu }{216}+\frac{25}{432}\right)+\frac{54}{35} \left(\frac{19 \nu ^2}{27}-\frac{25 \nu }{324}-\frac{17}{324}\right)\right]$ 
\\
$u^9$ & $\frac{1}{10} \ch \left(\ch^2-2\right) (2 \nu +1)^2 \sh^2 v$  
\\
\hline
Coefficient& $C_{K_1{\rm sin}}^{\rm 2PN, NNLO}$ \\
\hline
$u^4$ & $\sh \left[\frac{-\frac{1226 \nu ^2}{315}+\frac{4022 \nu }{315}+\frac{524}{315}}{\ch^4}+\ch^2 \left(\frac{968 \nu ^2}{105}-\frac{2978 \nu}{105}-\frac{248}{15}\right)+\frac{\frac{1226 \nu ^2}{315}-\frac{7478 \nu }{315}+\frac{5668}{315}}{\ch^2}+\frac{92 \nu ^2}{45}-\frac{3242 \nu}{315}+\frac{2804}{63}\right]$
\\
& $+A \left[\frac{8}{35} \ch^2 (102 \nu -391)-\frac{864 \nu }{35}+\frac{4432}{35}\right]$
\\
$u^5$ & $\pi \ch\sh \left[\frac{-\frac{1226 \nu ^2}{945}+\frac{4022 \nu }{945}+\frac{524}{945}}{\ch^4}-\frac{56 \nu ^2}{45}-\frac{2029 \nu}{315}-\frac{286}{45}+\frac{ \frac{344}{35}-\frac{192 \nu }{35}}{\ch^2}\right]
+\pi \ch  A \left[\frac{456}{35} \ch^2 \left(\nu +\frac{97}{38}\right)-\frac{432}{35} \left(\nu +\frac{311}{54}\right)\right]$
\\
$u^6$ & $\sh v^2 \left[\ch^2 \left(\frac{8 \nu ^2}{3}+\frac{1826 \nu }{105}+\frac{3932}{105}\right)+\frac{\frac{613 \nu ^2}{945}-\frac{2011 \nu}{945}-\frac{262}{945}}{\ch^2}+\frac{29 \nu ^2}{945}-\frac{12041 \nu }{945}+\frac{94678}{945}\right]$
\\
 & $+v \left[\ch^3 \left(\frac{62 \nu ^2}{7}+\frac{662 \nu }{35}+\frac{109}{5}\right)+\frac{-\frac{2452 \nu ^2}{945}+\frac{8044 \nu}{945}+\frac{1048}{945}}{\ch^3}+\ch \left(-\frac{6451 \nu ^2}{945}-\frac{2147 \nu }{135}+\frac{17863}{945}\right)+\frac{\frac{1226 \nu ^2}{945}-\frac{2878 \nu}{189}+\frac{18052}{945}}{\ch}\right]$
\\
 & $+\pi ^2 \sh \left[\ch^2 \left(-\frac{172 \nu ^2}{105}+\frac{163 \nu }{105}-\frac{29}{105}\right)+\frac{-\frac{613 \nu ^2}{3780}+\frac{2011 \nu
   }{3780}+\frac{131}{1890}}{\ch^2}-\frac{677 \nu ^2}{3780}+\frac{115 \nu }{756}-\frac{683}{1890}\right]$
\\
 & $+\sh \left[-\frac{8}{5} \ch^4 \left(\nu+\frac{1}{2}\right)^2+\ch^2 \left(-\nu ^2-\frac{279 \nu }{35}+\frac{1}{35}\right)-\frac{27 \nu ^2}{5}+\frac{3709 \nu }{105}-\frac{1591}{105}\right]$
\\
 & $+A \left[\frac{1472}{105} \ch^4 \left(\nu +\frac{83}{46}\right)-\frac{16}{5} \ch^2 \left(\nu -\frac{17}{7}\right)-\frac{2048 \nu }{35}-\frac{8}{35}+\ch\sh v \left(\frac{416}{35} \ch^2 \left(\nu +\frac{25}{26}\right)+\frac{64}{105} (\nu+410)\right)\right]$
\\
$u^7$ & $\pi  \ch \left(-\frac{649 \nu^2}{630}-\frac{4709 \nu }{630}-\frac{6358}{315}\right) \sh v^2
+\pi  A \ch\left[\frac{1024}{35}  \left(\nu+\frac{1}{256}\right)-\frac{2104}{105} \ch^2 \left(\nu+\frac{155}{526}\right)\right]$
\\
 & $+\pi v \left[ \ch^4 \left(-\frac{164 \nu ^2}{35}-\frac{242 \nu }{35}-\frac{23}{7}\right)+\ch^2 \left(\frac{67 \nu ^2}{30}-\frac{2581 \nu}{210}+\frac{523}{70}\right)+\frac{-\frac{613 \nu ^2}{945}+\frac{2011 \nu }{945}+\frac{262}{945}}{\ch^2}+\frac{319 \nu ^2}{210}+\frac{2431\nu }{210}-\frac{421}{15}\right]$
\\
 & $+\pi \ch \sh \left[ \ch^2 \left(-\frac{274 \nu ^2}{35}+\frac{99 \nu }{35}-\frac{79}{70}\right)+\frac{29 \nu ^2}{5}-\frac{112 \nu
   }{15}-\frac{7}{30}+\pi ^2 \left(\frac{353 \nu ^2}{1512}-\frac{1633 \nu }{7560}+\frac{47}{1890}\right)\right]$
\\
$u^8$ & $\ch v^3 \left[\ch^2 \left(-\frac{148 \nu ^2}{315}-\frac{241 \nu }{105}+\frac{1243}{90}\right)+\frac{484 \nu ^2}{945}+\frac{430 \nu}{189}-\frac{30946}{945}\right]
+\sh v^2 \left[-\frac{4}{5} \ch^4 \left(\nu +\frac{1}{2}\right)^2+\ch^2 \left(-\frac{121 \nu ^2}{70}-\frac{227 \nu}{42}-\frac{419}{210}\right)\right.$
\\
 & $\left.-\frac{3 \nu ^2}{35}-\frac{1601 \nu }{105}-\frac{1066}{105}\right]
+\ch v \left[-\frac{1}{5} \sh^2 \left(3 \ch^2-1\right) (2 \nu +1)^2+\pi ^2 \left(\ch^2\left(\frac{4 \nu ^2}{15}+\frac{13 \nu }{35}+\frac{407}{840}\right)-\frac{31 \nu ^2}{315}+\frac{82 \nu }{315}+\frac{137}{630}\right)\right]$
\\
 & $+\pi ^2 \sh \left[\frac{1}{5} \ch^4 \left(\nu +\frac{1}{2}\right)^2+\frac{177}{280} \ch^2 \left(\nu ^2-\frac{293 \nu }{531}-\frac{337}{531}\right)+\frac{59 \nu^2}{140}+\frac{173 \nu }{420}+\frac{31}{42}\right]
-\frac{8}{7} A \ch \left(\ch^2-4\right) (4 \nu -1) \sh v$
\\
$u^9$ & $\pi  \ch\sh v^2 \left[-\frac{27}{35} \ch^2 \left(\nu ^2+\frac{73 \nu }{18}+\frac{227}{108}\right)-\frac{27}{35}  \left(-\frac{16 \nu ^2}{9}-\frac{689 \nu}{81}-\frac{361}{81}\right)\right]+\frac{1}{10} \pi  \left(\ch^2+2\right) (2 \nu +1)^2 \sh^2 v$
\\
 & $+\pi ^3 \ch\sh \left[-\frac{27}{35} \ch^2 \left(-\frac{41 \nu ^2}{324}+\frac{19 \nu }{648}+\frac{25}{1296}\right)-\frac{27}{35}  \left(\frac{19 \nu^2}{81}-\frac{25 \nu }{972}-\frac{17}{972}\right)\right]$
\\
$u^{10}$ & $\frac{1}{120} \ch \left(\ch^2-2\right) (2 \nu +1)^2 \sh^2 v \left(3 \pi ^2-4 v^2\right)$
\\
\end{tabular}
\end{ruledtabular}
\end{table*}


\begin{table*}
\caption{\label{2PNcoeffs6}
2PN $C_{K_\mu\partial^2K_\nu}$, $C_{\partial^2K_\nu{\rm cos}}$ and $C_{\partial^2K_\nu{\rm sin}}$ coefficients.
}
\begin{ruledtabular}
\begin{tabular}{c|c|c|c|c}
Coefficient & $C_{K_0\partial^2K_0}^{\rm 2PN, NNLO}$ & $C_{K_1\partial^2K_0}^{\rm 2PN, NNLO}$ & $C_{K_0\partial^2K_1}^{\rm 2PN, NNLO}$ & $C_{K_1\partial^2K_1}^{\rm 2PN, NNLO}$ \\
\hline
$u^4$ & $-\frac{128 \nu ^2}{315}+\frac{160 \nu }{21}-\frac{1024}{15}$ & - & - & $\frac{128 \nu ^2}{105}-\frac{2192 \nu }{105}+\frac{5336}{63}$\\
$u^5$ & - & $-\frac{32 \nu^2}{27}-\frac{2384 \nu }{315}+\frac{71864}{945}$ & $\frac{1184 \nu^2}{945}-\frac{2384 \nu }{315}-\frac{22952}{189}$ & $$\\
$u^6$ & $-\frac{928 \nu^2}{945}-\frac{15608 \nu }{315}-\frac{176}{27}$ & - & - & $-\frac{1856 \nu ^2}{945}-\frac{15040 \nu }{189}+\frac{33232}{2835}$\\
$u^7$ & - & $\frac{32 \nu ^2}{945}-\frac{33136 \nu }{945}+\frac{4696}{81}$ & $\frac{32 \nu ^2}{945}-\frac{33136 \nu }{945}+\frac{4696}{81}$ & $$\\
$u^8$ & $-\frac{8 \nu ^2}{105}-\frac{30992 \nu }{945}+\frac{12616}{2835}$ & - & - & $\frac{2008 \nu ^2}{945}-\frac{27952 \nu }{945}+\frac{9896}{2835}$\\
$u^9$ & - & $\frac{464 \nu ^2}{189}+\frac{64 \nu }{21}-\frac{176}{189}$ & $\frac{464 \nu ^2}{189}+\frac{64 \nu }{21}-\frac{176}{189}$ & $$\\
$u^{10}$ & $-\frac{32 \nu ^2}{63}+\frac{64 \nu }{189}-\frac{32}{567}$ & - & - & $-\frac{32 \nu ^2}{63}+\frac{64 \nu }{189}-\frac{32}{567}$\\
\hline
Coefficient& \multicolumn{2}{c|}{$C_{\partial^2K_0{\rm cos}}^{\rm 2PN, NNLO}$}& \multicolumn{2}{c}{$C_{\partial^2K_0{\rm sin}}^{\rm 2PN, NNLO}$} \\
\hline
$u^6$ & \multicolumn{2}{c|}{
$\ch^2 \left(-\frac{22 \nu ^2}{945}+\frac{410 \nu }{189}+\frac{508}{945}\right)+\frac{263\nu ^2}{945}-\frac{79 \nu }{189}+\frac{176}{189}$
}& \multicolumn{2}{c}{
-
}\\
& \multicolumn{2}{c|}{
$+\frac{\frac{613 \nu ^2}{1890}-\frac{2011 \nu }{1890}-\frac{131}{945}}{\ch^2}$
}& \multicolumn{2}{c}{}
\\
$u^7$ & \multicolumn{2}{c|}{
$\pi  \ch^3 \left(\frac{191 \nu ^2}{126}-\frac{811 \nu }{630}+\frac{58}{315}\right)+\pi  \ch \left(-\frac{33 \nu ^2}{28}+\frac{491 \nu }{420}-\frac{13}{105}\right)$
}& \multicolumn{2}{c}{
$v \left(\ch^3 \left(\frac{688 \nu ^2}{315}+\frac{1184 \nu }{315}+\frac{323}{315}\right)+\ch \left(-\frac{52 \nu ^2}{35}-\frac{24 \nu }{7}-\frac{73}{30}\right)\right)$
}\\
& \multicolumn{2}{c|}{}& \multicolumn{2}{c}{
$+\sh \left[\ch^2 \left(\frac{221 \nu ^2}{35}-\frac{557 \nu }{105}+\frac{32}{15}\right)+\frac{\frac{613 \nu ^2}{945}-\frac{2011 \nu}{945}-\frac{262}{945}}{\ch^2}\right.$
}\\
& \multicolumn{2}{c|}{}& \multicolumn{2}{c}{
$\left.-\frac{563 \nu ^2}{1890}-\frac{3391 \nu }{1890}-\frac{319}{1890}\right]$
}\\
$u^8$ & \multicolumn{2}{c|}{
$\sh v \left(\ch^3 \left(\frac{64 \nu ^2}{35}+\frac{256 \nu }{105}+\frac{23}{21}\right)+\ch \left(\frac{76 \nu ^2}{45}+\frac{1562 \nu}{315}+\frac{1579}{630}\right)\right)$
}& \multicolumn{2}{c}{
$\pi  \sh \left(\ch^3 \left(-\frac{22 \nu ^2}{21}-\frac{3 \nu }{35}+\frac{11}{210}\right)+\ch \left(-\frac{1027 \nu ^2}{630}+\frac{836 \nu}{315}-\frac{391}{1260}\right)\right)$
}\\
& \multicolumn{2}{c|}{
$+\sh^2 \left(\frac{192}{35} \ch^2 \left(\nu -\frac{11}{48}\right) \left(\nu +\frac{1}{4}\right)+\frac{73 \nu ^2}{35}+\frac{43 \nu }{21}+\frac{641}{210}\right)$
}& \multicolumn{2}{c}{
$$
}\\
$u^9$ & \multicolumn{2}{c|}{
$\pi  \sh^2 \left(-\frac{2}{15} \ch^3 \left(\nu +\frac{1}{2}\right)^2-\frac{2}{15} \ch \left(\frac{249 \nu ^2}{14}-\frac{2 \nu }{7}-\frac{47}{56}\right)\right)$
}& \multicolumn{2}{c}{
$\sh^2 v \left(\ch \left(-\frac{138 \nu ^2}{35}-\frac{706 \nu}{105}-\frac{709}{210}\right)-\frac{4}{15} \ch^3 \left(\nu+\frac{1}{2}\right)^2\right)$
}\\
& \multicolumn{2}{c|}{}& \multicolumn{2}{c}{
$-\frac{1}{5} \left(\ch^2+1\right) (2 \nu+1)^2 \sh^3$
}\\
$u^{10}$ & \multicolumn{2}{c|}{
$-\frac{4}{5} \ch \left(\nu +\frac{1}{2}\right)^2 \sh^3 v$
}& \multicolumn{2}{c}{
$\frac{2}{5} \pi\ch \left(\nu +\frac{1}{2}\right)^2 \sh^3$
}\\
\hline
Coefficient& \multicolumn{2}{c|}{$C_{\partial^2K_1{\rm cos}}^{\rm 2PN, NNLO}$}& \multicolumn{2}{c}{$C_{\partial^2K_1{\rm sin}}^{\rm 2PN, NNLO}$} \\
\hline
$u^6$ & \multicolumn{2}{c|}{
-
}& \multicolumn{2}{c}{
$\sh \left[\ch^2 \left(\frac{688 \nu ^2}{105}-\frac{652 \nu }{105}+\frac{116}{105}\right)+\frac{\frac{613 \nu ^2}{945}-\frac{2011 \nu}{945}-\frac{262}{945}}{\ch^2}\right.$
}\\
& \multicolumn{2}{c|}{}& \multicolumn{2}{c}{
$\left.+\frac{677 \nu ^2}{945}-\frac{115 \nu }{189}+\frac{1366}{945}\right]$
}\\
$u^7$ & \multicolumn{2}{c|}{
$\ch \left(\frac{1144 \nu ^2}{315}+\frac{2672 \nu }{315}\frac{1349}{315}\right) \sh v$
}& \multicolumn{2}{c}{
$\pi  \sh \left(-\frac{353 \ch \nu ^2}{126}+\frac{1633 \ch \nu }{630}-\frac{94 \ch}{315}\right)$
}\\
& \multicolumn{2}{c|}{
$+\ch^4 \left(\frac{192 \nu ^2}{35}+\frac{4 \nu }{35}-\frac{11}{35}\right)+\ch^2 \left(-\frac{128 \nu ^2}{105}+\frac{128 \nu }{35}+\frac{32}{7}\right)$
}& \multicolumn{2}{c}{}
\\
& \multicolumn{2}{c|}{
$+\frac{\frac{613\nu ^2}{945}-\frac{2011 \nu }{945}-\frac{262}{945}}{\ch^2}-\frac{373\nu ^2}{210}-\frac{97 \nu }{42}-\frac{158}{105}$
}& \multicolumn{2}{c}{}
\\
$u^8$ & \multicolumn{2}{c|}{
$\pi  \left(\ch^3 \left(-\frac{25 \nu ^2}{84}-\frac{293 \nu }{140}+\frac{71}{210}\right)+\ch \left(-\frac{7 \nu ^2}{18}+\frac{367 \nu}{126}-\frac{247}{630}\right)\right)$
}& \multicolumn{2}{c}{
$v \left(\ch^3 \left(-\frac{16 \nu ^2}{15}-\frac{52 \nu }{35}-\frac{407}{210}\right)+\ch \left(\frac{124 \nu ^2}{315}-\frac{328 \nu}{315}-\frac{274}{315}\right)\right)$
}\\
& \multicolumn{2}{c|}{}& \multicolumn{2}{c}{
$+\sh \left[-\frac{4}{5} \ch^4 \left(\nu +\frac{1}{2}\right)^2+\ch^2 \left(-\frac{177 \nu ^2}{70}+\frac{293 \nu }{210}+\frac{337}{210}\right)\right.$
}\\
& \multicolumn{2}{c|}{}& \multicolumn{2}{c}{
$\left.-\frac{59 \nu ^2}{35}-\frac{173 \nu }{105}-\frac{62}{21}\right]$
}\\
$u^9$ & \multicolumn{2}{c|}{
$\sh v \left(\ch^3 \left(\frac{68 \nu ^2}{35}+\frac{20 \nu }{7}+\frac{101}{70}\right)+\ch \left(-\frac{124 \nu ^2}{35}-\frac{664 \nu}{105}-\frac{344}{105}\right)\right)$
}& \multicolumn{2}{c}{
$\pi  \sh \left(\ch^3 \left(-\frac{41 \nu ^2}{35}+\frac{19 \nu }{70}+\frac{5}{28}\right)+\ch \left(\frac{76 \nu ^2}{35}-\frac{5 \nu}{21}-\frac{17}{105}\right)\right)$
}\\
& \multicolumn{2}{c|}{
$-\frac{1}{10} \left(\ch^2+2\right) (2 \nu +1)^2 \sh^2$
}& \multicolumn{2}{c}{}
\\
$u^{10}$ & \multicolumn{2}{c|}{
$-\frac{1}{20} \pi  \ch \left(\ch^2-2\right) (2 \nu +1)^2 \sh^2$
}& \multicolumn{2}{c}{
$-\frac{1}{10} \ch \left(\ch^2-2\right) (2 \nu +1)^2 \sh^2 v$
}\\
\end{tabular}
\end{ruledtabular}
\end{table*}

\section{Emission of electromagnetic dipole radiation by two charged particles in hyperbolic motion}
\label{LL}

Let us recall the Landau-Lifshitz derivation of the dipole radiation due to the Coulombian interaction of two attracting charges $e_1$ and $e_2$ ($e_1e_2<0$).
The motion is assumed to be nonrelativistic and hyperbolic.
The dipole moment of the system writes
\beq
{\mathbf d}=e_1 {\mathbf  r}_1 +e_2 {\mathbf  r}_2=(e_1X_2-e_2X_1){\mathbf r}\,,
\eeq
since ${\mathbf  r}_1=X_2 {\mathbf r}$ and ${\mathbf  r}_2=-X_1 {\mathbf r}$, with $X_{1,2}=\frac{\mu}{m_{2,1}}$ and $ {\mathbf  r}={\mathbf  r}_1-{\mathbf  r}_2$ the relative position.
The relative motion corresponds to that of particle with mass the reduced mass $\mu$ along the hyperbola with parametric equations
\bea
\bar n t &=& e_r\sinh v-v\,,\nonumber\\
r&=& \bar a_r (e_r\cosh v-1)\,,\nonumber\\
\phi&=& 2\,{\rm arctan}\left[\sqrt{\frac{e_r+1}{e_r-1}}\tanh \frac{v}{2}\right]\,,
\eea
where $v\in (-\infty,\infty)$ and
\beq
\bar n = \sqrt{\frac{\alpha}{\mu \bar a_r^3}}
\,,\qquad
\bar a_r=\frac{\alpha}{2{\mathcal E}}
\,,\qquad
e_r=\sqrt{1+\frac{2{\mathcal E}J^2}{\mu\alpha^2}}
\,,
\eeq
with $\alpha=|e_1e_2|$.
Here ${\mathcal E}=E_{\rm tot}-Mc^2>0$, and dimensions are such that
\beq
[\bar a_r]=[L]\,,\quad [\alpha]=[e_1^2]=[cJ]=[Mc^2][ L]\,.
\eeq

The dipole radiation emitted during the whole scattering process is given by
\bea
\label{deltaEemdip}
\Delta E_{\rm em}^{\rm dip}&=&\frac{2}{3c^3}\int_{-\infty}^\infty dt (\ddot {\mathbf d})^2\nonumber\\
&=&\frac{2}{3c^3}(e_1X_2-e_2X_1)^2\int_{-\infty}^\infty dt (\ddot{\mathbf r})^2\nonumber\\
&=&\frac{2}{3c^3}(e_1X_2-e_2X_1)^2\int_{-\infty}^\infty dt (\ddot x^2+\ddot y^2)\,,
\eea
where the overdot denotes a time derivative, and ${\mathbf r}=x(t) \partial_x +y(t) \partial_y$.

Substituting the Fourier transforms 
\bea
x(t)&=&\int_{-\infty}^\infty \frac{d\omega}{2\pi}e^{-i\omega t}\hat x(\omega)
\,, \nonumber\\
y(t)&=&\int_{-\infty}^\infty \frac{d\omega}{2\pi}e^{-i\omega t}\hat y(\omega)
\,,
\eea
into the above integral leads to
\bea
\label{deltaEemdipom}
\Delta E_{\rm em}^{\rm dip}&=&\frac{4}{3c^3}(e_1X_2-e_2X_1)^2\nonumber\\
&&\times
\int_0^\infty \frac{d\omega}{2\pi} \omega^4 \left(|\hat x(\omega)|^2+|\hat y(\omega)|^2\right)\,,
\eea
where
\bea
\hat x(\omega)&=&\int_{-\infty}^\infty dt e^{i\omega t}x(t)
\,, \nonumber\\
\hat y(\omega)&=&\int_{-\infty}^\infty dt e^{i\omega t}y(t)
\,.
\eea
Using our parametrization in terms of the variable $v$, i.e., 
\bea
x&=&r\cos\phi=\bar a_r(e_r-\cosh v)\,,\nonumber\\
y&=&r\sin\phi=\bar a_r \sqrt{e_r^2-1}\sinh v\,,
\eea
the previous Fourier components become
\bea
\label{hatxom}
\hat x(\omega)&=&\int_{-\infty}^\infty \frac{dt}{dv} e^{i\omega t(v)}x(t(v)) dv\nonumber\\
&=&
\frac{\bar a_r}{\bar n}\int_{-\infty}^\infty dv\, e^{i\frac{\omega}{\bar n}(e_r\sinh v-v)}\nonumber\\
&&\times 
(e_r\cosh v-1) (e_r-\cosh v)  \nonumber\\
&=&\frac{  \pi \bar a_r}{\omega }H_{p}^{(1)}{}'(q)\,,
\eea
with
\beq
p=\frac{q}{e_r}\,,\qquad
q=iu\,,\qquad
u=\frac{\omega}{\bar n}e_r\,,
\eeq
and 
\bea
\label{hatyom}
\hat y(\omega)&=&\frac{\bar a_r\sqrt{e_r^2-1}}{\bar n}\int_{-\infty}^\infty dv\, e^{i\frac{\omega}{\bar n}(e_r\sinh v-v)}\nonumber\\
&&\times
\sinh v (e_r-\cosh v) \nonumber\\
&=&-\frac{\pi \bar a_r\sqrt{e_r^2-1}}{\omega e_r }H_{p}^{(1)}(q)\,,
\eea
respectively, where we have used the representation \eqref{Hankel_rep} for the Hankel function of the first kind as well as the standard identities between Bessel functions.

Finally, the dipole radiation \eqref{deltaEemdipom} reads (see Eq. (70.18) in Ref. \cite{Landau:1982dva})
\bea
\Delta E_{\rm em}^{\rm dip}&=&\frac{2\pi}{3c^3}(e_1X_2-e_2X_1)^2\bar a_r^2\nonumber\\
&\times&
\int_0^\infty d\omega\, \omega^2 \left([H_{p}^{(1)}{}'(q)]^2
+\frac{e_r^2-1}{e_r^2}[H_{p}^{(1)}(q)]^2\right)\,.\nonumber\\
\eea
The integration over the frequencies cannot be performed in closed analytical form by using standard methods, because the order $p$ of the Hankel functions also depend on $\omega$.
Landau and Lifshitz then studied the limiting cases of low and high frequencies, for which asymptotic expressions of the Hankel functions are available.

Let us apply to the above integral the same procedure as for the gravitational wave energy, by taking the large-eccentricity limit of the integrand.
First of all, it is convenient to replace the integration variable $\omega$ by $u$ in Eq. \eqref{deltaEemdipom}, yielding

\begin{widetext}

\bea
\label{deltaEemdipu}
\Delta E_{\rm em}^{\rm dip}&=&\frac{2}{3\pi c^3}(e_1X_2-e_2X_1)^2\frac{\bar n^5}{e_r^5}\int_{0}^\infty du\, u^4 \left(|\hat x(u)|^2+|\hat y(u)|^2\right)\nonumber\\
&=&\frac{8}{3\pi c^2}(e_1X_2-e_2X_1)^2\frac{\bar a_r^2\bar n^3}{e_r^3}\int_{0}^\infty du\,{\mathcal H}(u)
\,,
\eea
where 
\beq
{\mathcal H}(u)=u^2 e^{\pi \frac{u}{e_r}} \left[K_{1+p}^2(u)-\frac{2i}{e_r}K_{1+p}(u)K_{p}(u)+\left(1 -\frac{2}{e_r^2}\right) K_p^2(u)\right]\,,
\eeq
and we have used the equivalent expressions
\bea
\hat x(u)
&=&\frac{ 2 \bar a_r e_r}{u \bar n}e^{\frac{\pi}{2}\frac{u}{e_r}} \left(K_{p+1}(u) - \frac{i}{e_r} K_p(u)\right)\nonumber\\
\hat y(u)
&=&  i\frac{2 \bar a_r\sqrt{e_r^2-1}}{u \bar n} e^{\frac{\pi}{2}\frac{u}{e_r} }K_p(u)
\,.
\eea
for the Fourier components \eqref{hatxom} and \eqref{hatyom} in terms of the modified Bessel functions of the first kind (see Eq. \eqref{H1vsK}).
Next, expand the integrand ${\mathcal H}(u)$ in power series of the large eccentricity up to the NNLO, i.e., 
\bea
{\mathcal H}(u)&=&{\mathcal H}^{\rm LO}(u)+\frac{\pi}{e_r}{\mathcal H}^{\rm NLO}(u)+\frac{1}{e_r^2}{\mathcal H}^{\rm NNLO}(u)
+O\left(\frac{1}{e_r^3}\right)\,.
\eea
We find
\bea
{\mathcal H}^{\rm LO}(u)&=&u^2\left[K_{0}^2(u)+K_1^2(u)\right]\nonumber\\
&=&u^2[f_{K_0K_0}(u)+f_{K_1K_1}(u)]
\,,\nonumber\\
{\mathcal H}^{\rm NLO}(u)&=&u{\mathcal H}^{\rm LO}(u)\nonumber\\
&=&u^3[f_{K_0K_0}(u)+f_{K_1K_1}(u)]
\,,\nonumber\\
{\mathcal H}^{\rm NNLO}(u)&=&\frac{\pi^2u^2}{2}{\mathcal H}^{\rm LO}(u)-u^2K_{0}^2(u)
-u^4\left[K_{0}(u)\frac{\partial^2 K_\nu(u)}{\partial \nu^2 }\Bigg|_{\nu=0}+K_{1}(u)\frac{\partial^2 K_\nu(u)}{\partial \nu^2 }\Bigg|_{\nu=1}\right]\nonumber\\
&=&\left(\frac{\pi^2u^4}{2}-u^2\right)f_{K_0K_0}(u)+\frac{\pi^2u^4}{2}f_{K_1K_1}(u)-u^4[f_{K_0\partial^2K_0}(u)+f_{K_1\partial^2K_1}(u)]
\,.
\eea
Integration over $u$ is done by taking the Mellin transform
\bea
\int_0^\infty du\,{\mathcal H}^{\rm LO}(u)&=&g_{K_0K_0}(3)+g_{K_1K_1}(3)
=\frac{\pi^2}{8}
\,,\nonumber\\
\int_0^\infty du\,{\mathcal H}^{\rm NLO}(u)&=&g_{K_0K_0}(4)+g_{K_1K_1}(4)
=1
\,,\nonumber\\
\int_0^\infty du\,{\mathcal H}^{\rm NNLO}(u)&=&\frac{\pi^2}{2}g_{K_0K_0}(5)-g_{K_0K_0}(3)+\frac{\pi^2}{2}g_{K_1K_1}(5)-g_{K_0\partial^2K_0}(5)-g_{K_1\partial^2K_1}(5)
=\frac{9\pi^2}{16}
\,.
\eea
The averaged dipole radiation \eqref{deltaEemdipu} thus reads
\bea
\Delta E_{\rm em}^{\rm dip}&=&\frac{8}{3 c^2}(e_1X_2-e_2X_1)^2\frac{\bar a_r^2\bar n^3}{e_r^3}
\left[\frac{\pi}{8} +\frac{1}{e_r}+ \frac{9\pi}{16e_r^2}+O\left(\frac{1}{e_r^3}\right)\right]
\,.
\eea
\end{widetext}
It is straightforward to proceed further in the expansion.

One can easily check this result by performing the integral \eqref{deltaEemdip} directly in the time domain.
In fact, passing to the variable $v$ we get
\bea
\label{deltaEemdipexact}
\Delta E_{\rm em}^{\rm dip}&=&\frac{2}{3c^3}(e_1X_2-e_2X_1)^2\bar a_r^2 \bar n^3 \int_{-\infty}^\infty\frac{dv}{(e_r\cosh v -1)^3}\nonumber\\
&=& \frac{2}{3c^3}(e_1X_2-e_2X_1)^2\bar a_r^2 \bar n^3 f(e_r)\,,
\eea
where
\bea
f(e_r)&=&\int_{-\infty}^\infty\frac{dv}{(e_r\cosh v -1)^3}\nonumber\\
&=&\frac{1}{(e_r^2-1)^2}\left[3+\frac{e_r^2+2}{\sqrt{e_r^2-1}}{\rm arccos}\left(-\frac{1}{e_r}\right)\right]\nonumber\\
&=&\frac{\pi}{2 e_r^3} +\frac{4}{e_r^4}+ \frac{9\pi}{4e_r^5}+ \frac{32}{3e_r^6}+O \left(\frac{1}{e_r^7}\right)\,.
\eea
This is also in agreement with the expression for $\Delta E_{\rm em}^{\rm dip}$ in terms of the deflection angle $\chi$ given in the Problem 2 at the end of the paragraph 70 in Ref. \cite{Landau:1982dva}.
It is enough to substitute in it the relation $\cot\frac{\chi}{2}=b\mu c^2 p_\infty^2/\alpha$ (or equivalently $\frac{\chi}{2}=\arccos(-\frac{1}{e_r})-\frac{\pi}{2}$) between the deflection angle and the impact parameter $b$, recalling that ${\mathcal E}=\mu c^2 p_\infty^2/2$, $J=b\mu c p_\infty$, $\bar a_r^2 \bar n^3=\mu c^5 p_\infty^5/\alpha$ and $b=\alpha\sqrt{e_r^2-1}/(c p_\infty)$.

Landau and Lifshitz also computed the total \lq\lq effective radiation'' emitted during the scattering of a parallel beam of particles (Problem 3) by multiplying the total dipolar energy $\Delta E_{\rm em}^{\rm dip}$ by $2\pi bdb$ and integrating over all values of the impact parameter $b\in[0,\infty)$, i.e.,
\beq
\varkappa_{\rm em}^{\rm dip}=2\pi\int_0^\infty \Delta E_{\rm em}^{\rm dip}\,b\,db\,.
\eeq
Substituting the exact result \eqref{deltaEemdipexact} for $\Delta E_{\rm em}^{\rm dip}$ with $e_r$ replaced by its expression in terms of $b$ and taking the finite part finally give
\beq
\varkappa_{\rm em}^{\rm dip}
=\frac{8\pi}{9c^2}\frac{\alpha p_\infty}{\mu}(e_1X_2-e_2X_1)^2\,.
\eeq


\end{document}